\documentclass[reprint,amsmath,amssymb,aps,longbibliography,letterpaper,pra]{revtex4-1}

\usepackage[dvipsnames]{xcolor}
\usepackage[normalem, normalbf]{ulem}
\usepackage{graphicx}
\usepackage{dcolumn}
\usepackage{bm}
\usepackage{amsmath, amsthm,amssymb}
\usepackage[colorlinks=true, citecolor=Blue,
            linkcolor=BrickRed, urlcolor=ForestGreen]{hyperref}
\usepackage{changes}
\usepackage{txfonts}
\usepackage{epstopdf}
\usepackage{verbatim}
\usepackage{appendix}

\newcommand*{\affmark}[1][*]{\textsuperscript{#1}}
\begin{document}

\title{Gravitational wave detectors with broadband high frequency sensitivity}
\author{Michael A. Page\affmark[1], Maxim Goryachev\affmark[2], Haixing Miao\affmark[3], Yanbei Chen\affmark[4], Yiqiu Ma\affmark[5], David Mason\affmark[6], Massimiliano Rossi\affmark[7], Carl D. Blair\affmark[1], Li Ju\affmark[1], David G. Blair\affmark[1], Albert Schliesser\affmark[7], Michael E. Tobar\affmark[2], and Chunnong Zhao\affmark[1]}

\affiliation{\affmark[1]Australian Research Council Center of Excellence for Gravitational Wave Discovery, University of Western Australia, 35 Stirling Highway, Perth, Western Australia 6009, Australia}
\affiliation{\affmark[2] Australian Research Council Center of Excellence for Engineered Quantum Systems, University of Western Australia, 35 Stirling Highway, Perth, Western Australia 6009, Australia}
\affiliation{\affmark[3] Astrophysics and Space Research Group, University of Birmingham, Birmingham B15 2TT, United Kingdom}
\affiliation{\affmark[4] Theoretical Astrophysics, California Institute of Technology, 1200 E California Blvd, Pasadena, California 91125, United States}
\affiliation{\affmark[5] Center for Gravitational Experiment,  School of Physics, Huazhong University of Science and Technology, Wuhan, 430074, China}
\affiliation{\affmark[6] Yale Quantum Institute, Yale University, 17 Hillhouse Ave, New Haven, Connecticut 06511, United States
}
\affiliation{\affmark[7] Niels Bohr Institute, University of Copenhagen, Blegdamsvej 17, 2100 Copenhagen, Denmark}

\begin{abstract}
The binary neutron star coalescence GW170817 was observed by gravitational wave detectors during the inspiral phase but sensitivity in the 1-5 kHz band was insufficient to observe the expected nuclear matter signature of the merger itself, and the process of black hole formation. This provides strong motivation for improving 1--5 kHz sensitivity which is currently limited by photon shot noise. Resonant enhancement by signal recycling normally improves the signal to noise ratio at the expense of bandwidth. The concept of optomechanical white light signal recycling (WLSR) has been proposed, but all schemes to date have been reliant on the development of suitable ultra-low mechanical loss components. Here for the first time we show demonstrated optomechanical resonator structures that meet the loss requirements for a WLSR interferometer with strain sensitivity below 10$^{-24}$ Hz$^{-1/2}$ at a few kHz. Experimental data for two resonators are combined with analytic models of 4km interferometers similar to LIGO, to demonstrate sensitivity enhancement across a much broader band of neutron star coalescence frequencies than dual-recycled Fabry-Perot Michelson detectors of the same length. One candidate resonator is a silicon nitride membrane acoustically isolated from the environment by a phononic crystal. The other is a single-crystal quartz lens that supports bulk acoustic longitudinal waves. Optical power requirements could prefer the membrane resonator, although the bulk acoustic wave resonator gives somewhat better thermal noise performance. Both could be implemented as add-on components to existing detectors. 
\end{abstract}

\maketitle

Since the detection of gravitational waves (GW) from binary black holes and neutron stars \cite{GWDiscovery, NSDiscovery, O2Catalogue, SecondNS}, there is increasing interest in improving the sensitivity and bandwidth of detectors to allow better characterization of gravitational wave sources. Detectors such as the proposed Einstein Telescope \cite{punturo2010einstein} and Cosmic Explorer \cite{LSCCE} aim for improved low frequency sensitivity to dramatically increase the number of observable cycles from compact binary coalescence events. Other detectors focus on multimessenger astronomy from neutron star coalescences, targeting a strain sensitivity of $h \sim$ 10$^{-24}$ Hz$^{-1/2}$ in the 1--5 kHz band. Observation of the normal modes of new born hypermassive neutron stars will provide insight into the complex hydrodynamics of nuclear matter moments before its collapse into a black hole \cite{ClarkQNM}. Other sources of GWs in the range 1--5 kHz include the final moments of black hole coalescence, normal modes of new born black holes with mass 5--20 M$_{\rm \odot}$ and core collapse supernovae.

High frequency sensitivity in interferometric gravitational wave detectors is currently limited by quantum shot noise \cite{Aasi2013, BarsottiSqueezing} with strain sensitivity $h$ of a few times $10^{-23}$ Hz$^{1/2}$. A straightforward way to reduce the quantum shot noise level is to increase the laser power inside the detector. In addition, configurations based on detuning and strongly coupled signal recycling \cite{BuonnanoNoise} can produce a broadband response at high frequency, but achieving target sensitivity of $h \sim 10^{-24}$ Hz$^{1/2}$ still requires arm power levels an order of magnitude higher than the best attained to date.

 \begin{figure*}[!ht]
     \centering \includegraphics[width=\textwidth]{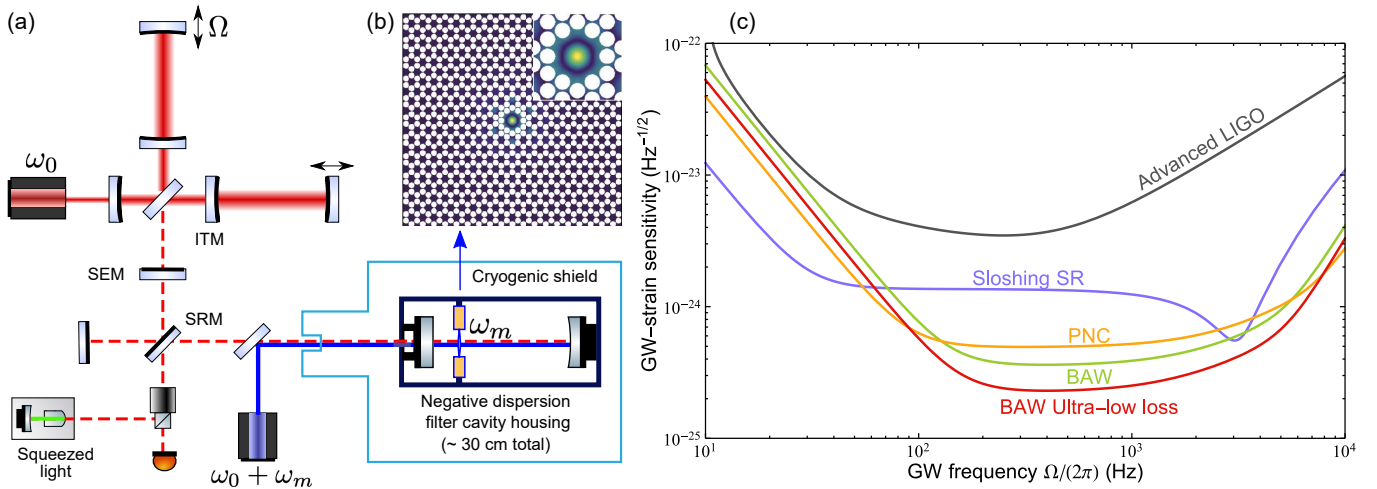}
     \caption{\textbf{WLSR interferometer configuration incorporating optomechanical negative dispersion:} \textbf{(a):} Interferometric GW detector with optomechanical filter coupled to the dark port signal recycling optics. GWs modulate the interferometer carrier, producing laser sidebands at $\omega_0 + \Omega$. The input test mass (ITM) and signal extraction mirror (SEM) are impedance matched to ensure maximum transmission of sidebands. The signal recycling mirror (SRM) couples the interferometer dark port, filter cavity and output photodetector. Squeezed vacuum is injected at the output Faraday isolator. The optomechanical negative dispersion filter cavity is pumped by blue detuned light at $\omega_0+\omega_m$. The cavity itself is 5 cm long but may be contained inside a larger housing. \textbf{(b):} Illustration of the PNC resonator, which consists of a silicon nitride membrane that functions as an effective defect in a phononic lattice. The colour scale represents the displacement of the out-of-plane mode of the resonator. \textbf{(c):} Quantum noise-limited strain sensitivity curves of various GW detectors. ``Advanced LIGO'' denotes the nominal design sensitivity of Advanced LIGO shown in \cite{ALIGODesign2015} at 800 kW arm cavity power. ``Sloshing SR'' refers to a detector where the length and transmission of the signal recycling cavity are tuned to achieve an optical resonance with a frequency of 2.5 kHz and bandwidth of 1.5 kHz. ``PNC'' refers to a WLSR interferometer using the setup shown in (a) and (b), and experimentally demonstrated values of mechanical quality factor for the PNC resonator \cite{MasonPNC}. ``BAW'' refers to a WLSR interferometer where the optomechanical component is a plano-convex bulk-acoustic wave resonator, and ``BAW Ultra-low loss'' is calculated for speculative improvements in BAW resonator quality factor, detailed in Supplementary Material. Apart from "Advanced LIGO", all curves use 4.0 MW arm cavity power and 10 dB frequency dependent squeezing.}
     \label{fig:layout}
 \end{figure*}

In general, signal recycling improves detector sensitivity by resonant enhancement of the signal rather than suppression of shot noise at the detection port. However, in conventional systems the resonance response creates a trade-off between sensitivity gain and bandwidth~\cite{MizunoSensitivity}. In principle, the sensitivity-bandwidth tradeoff can be overcome by the method of white light signal recycling (WLSR). While travelling across the long interferometer arms, the GW signal sidebands experience a phase delay relative to the carrier. A negative dispersion medium inside the signal recycling cavity can compensate for the signal sideband phase delay, creating a broadband resonance called a white light cavity~\cite{wicht1997white,zhou2015quantum}. The energetic quantum limit of the cavity is lowered via quantum amplification~\cite{braginsky2000energetic,BraginskySQL,tsang2011fundamental,miao2017towards}, indicating that the interferometer supports a non-classical state and physical laws are not violated. 

A succession of recent papers \cite{HaixingFilters, PageWLSR, HaixingHF, JoeTransmission} have shown that WLSR can be implemented by using an optomechanically coupled negative dispersion filter. The filter consists of a mechanical resonator placed inside a Fabry-Perot cavity with optical resonance $\omega_0$ equal to the interferometer carrier frequency. The cavity pump light has blue detuning equal to the mechanical resonance $\omega_m$, and is stabilized by feedback \cite{HaixingFilters}. The negative dispersion filter can be seen as a blue-detuned analogue of optomechanically-induced transparency, where the GW signals act as the near-resonant probe. Parametric interaction between the signal, pump light and mechanical resonator stores the signal with a frequency dependent phase compensation, creating the required negative dispersion effect.

To maintain quantum amplification, noises introduced by the mechanical resonator must be kept low. It has been shown that the mechanical resonator must have a quality factor $Q_m$ and operate at temperature $T$ such that $T/Q_{\rm m}$ $<$ 10$^{-9}$ K, in order for thermal noise not to dominate the detector noise budget \cite{HaixingFilters}. In addition, vacuum noise sidebands at $\omega_0 + 2\omega_m \pm \Omega$ are present inside the detuned cavity and create extra quantum noise at GW signal sideband frequencies. High $\omega_m$ is required to keep the extra sidebands far detuned from the interferometer resonance, and their impact can be further mitigated using a high finesse filter cavity \cite{JoeTransmission}. 

To date, the only proposed solution to the mechanical resonator thermal and quantum noise issues has been to use optical dilution to increase the resonant frequency and Q-factor of mechanically soft micro-pendulums \cite{YiqiuDilution, PageWLSR, HaixingHF}. However, optical dilution for the purpose of GW detection is technically demanding. The mechanical resonators need to be very small, and yet able to operate at high optical power densities. The trapping power required to achieve sufficient $Q_m$ results in coupling to other loss mechanisms, placing an upper limit on the viable $\omega_m$. Optical dilution must be balanced against thermoelastic loss, acceleration loss and beam size. This leaves a very small volume of parameter space in which the necessary performance might be achieved \cite{PageWLSR}, and until now a suitable optomechanical resonator has not been demonstrated.

In this paper we present the first schemes of optomechanical negative dispersion that have demonstrated levels of low mechanical loss suitable for broadband GW detectors. The first candidate is a silicon nitride membrane resonator isolated from the external environment by a phononic crystal (henceforth referred to as ``phononic crystal'' resonator or PNC). Mason, \textit{et al.} have maintained a $\omega_m/(2\pi)=1.135$ MHz out-of-plane vibrational mode at $Q_m = 1.03\times 10^9$ and $T=10$ K for a 20 nm thick Si$_3$N$_4$ membrane shielded with an acoustic bandgap of 1.07--1.28 MHz \cite{MasonPNC}. The phononic crystal can be optomechanically coupled by using it in a ``membrane-in-the-middle'' (MIM) configuration as characterised by Thompson \textit{et al.} \cite{ThompsonMIM}. Nanogram membrane resonators in MIM cavities have been shown to have strong optopmechanical coupling - for our design, we can achieve WLSR using filter cavity circulating power of 42.2 mW.

The second candidate is a plano-convex lens constructed from single crystal quartz, known as a Bulk Acoustic Wave (BAW) resonator due to its characteristic of bulk longitudinal phonons with extremely high quality factor. Galliou, \textit{et al.} have measured $\omega_m/(2\pi) = 204$ MHz and $Q_m = 8\times 10^9$ at 4~K for the 65th longitudinal mode of a 30 mm diameter, 1 mm thick quartz BAW resonator \cite{GalliouBAW}. Kharel, \textit{et al.} have demonstrated strong optomechanical coupling in BAW resonators using Brillouin scattering \cite{KharelBAW}. However, Brillouin scattering using near-infrared light requires a mechanical mode of approximately 18 GHz, which would have surface scattering losses that exceed the strict thermal noise requirements for WLSR \cite{GoryachevGeometry, GalliouBAW}. Optomechanical coupling to surface motion of the 204 MHz mode is possible in principle, and explored in Supplementary Material, but gives a low coupling rate. This in combination with the higher mass of the BAW resonator means that it requires much higher intracavity power to achieve WLSR, in excess of 10 kW, but the very low optical losses of quartz mean that the dissipated power could be manageable.

The WLSR interferometer layout is shown in figure \ref{fig:layout}. The negative dispersion filter is coupled to the signal recycling cavity. In the interferometer, the input test masses of the arm cavities are impedance matched to the signal extraction mirror, which allows for enhanced transmission of GW sidebands into the signal recycling cavity. The signal recycling mirror couples the interferometer dark port, negative dispersion filter and output photodiode. Frequency dependent squeezing may be applied by injecting squeezed vacuum at the output Faraday isolator \cite{KimbleSqueezing}. The negative dispersion filter is cryogenically cooled to liquid helium temperatures of 1--4 K and contained inside a radiation shield to minimise heating from external radiation and phase noise from scattered light. The PNC resonator is also shown in figure \ref{fig:layout}, where it is embedded in a 2-dimensional phononic lattice. High mechanical quality factors have been demonstrated for silicon nitride PNC resonators of size 87--346 $\mu$m \cite{DenmarkPNC}.
 
White light signal recycling using our candidate resonators is capable of producing a broader band of sensitivity enhancement compared to specialised high frequency dual-recycled Fabry-Perot Michelson detectors. For example, Martynov, \textit{et al.} showed that by tuning the transmissivity and length of the signal recycling cavity, an optical ``sloshing'' resonance at 2.5 kHz, with bandwidth 1.5 kHz, could be created in order to amplify neutron star signals \cite{DenisHF}. Figure \ref{fig:layout} compares our WLSR scheme with the sloshing resonance design, showing superior gain/bandwidth enhancement of quantum noise limited sensitivity in the 1--5 kHz range, at similar levels of interferometer optical loss, arm cavity power and optical squeezing. The WLSR interferometer has an additional advantage of being able to maintain a short signal recycling cavity of much less than 100 m. Nominal properties of the filter cavity and PNC resonator used to produce figure \ref{fig:layout} are shown in table \ref{tab:membraneMode}.

Throughout this paper we will discuss the optomechanics necessary to create broadband WLSR. We give an overview of the theory that leads to the key parameters of our negative dispersion filter design, particularly the circulating power. We then detail the main inputs into the quantum optical calculation used to produce the sensitivity spectrum shown in figure \ref{fig:layout}. Technical considerations that arise from integrating the negative dispersion filter into GW detectors are presented in the discussion section. The quantum optical framework of the sensitivity spectrum calculation is detailed in Methods. In the Supplementary Material we further elaborate on the optomechanics of the BAW resonator, the parameters used to produce the sensitivity curves of figure \ref{fig:layout}, the impact of interferometer optical losses and absorption heating of candidate resonators.

\begin{table}[]
    \centering
    \begin{tabular}{|c c c|}\hline
     Parameter    & Symbol & Value \\ \hline
        \textbf{Membrane resonator} & & \\
       Refractive index (Si$_3$N$_4$)  & $n_{\rm SiN}$ & 1.98\\
          Membrane thickness   & $h_m$ & 20 nm \\
          Mechanical frequency  &  $\omega_m/(2\pi)$  & $1.135$ MHz \\
         Mechanical Q-factor  &  $Q_m$  & 1.03$\times$10$^{9}$  \\
         Acoustic bandgap & & 1.07--1.28 MHz \\
         Effective mass & $M_{\rm eff}$ & 2.3 ng \\
         \textbf{Filter cavity} & & \\
          Length & $L_f$&0.05 m\\
                    Circulating power & $P_f$ & 42.2 mW \\
                 Input transmission & $T_f$ & 300 ppm \\
                 Temperature & $T$ & 1 K \\ \hline
                \end{tabular}
    \caption{Nominal properties of the negative dispersion filter cavity and PNC resonator discussed throughout this paper}
    \label{tab:membraneMode}
\end{table}

\section*{Optomechanical coupling in negative dispersion filters}\label{sec:power}

In an optomechanical cavity, the optomechanical coupling can be described as the energy changes per unit mechanical displacement. The associated interaction Hamiltonian can be written in a form that suggests correlated two-photon exchange \cite{HaixingFilters}:

\begin{equation}\label{eq:RWAinteraction}
    \hat{H}_{\rm int} = -\hbar g\left(\hat{a}\hat{b} + \hat{a}^{\dag}\hat{b}^{\dag}\right)
\end{equation}

\noindent
where $\hat{a}$ and $\hat{b}$ are the annihilation operators of the optical and mechanical modes inside the filter cavity, respectively, and $g$ is the optomechanical coupling rate. Using this Hamiltonian, the negative dispersion filter is shown to have the following input-output relation:
\begin{equation}\label{eq:inputOutput}
    \hat{a}_{\rm out}(\omega_0+\Omega) = \mathrm{Exp}\left[-2\mathbf{i}\Omega/\gamma_{\rm opt}\right] \hat{a}_{\rm in}(\omega_0+\Omega)
\end{equation}
\noindent
where $\hat{a}_{\rm in}$ and $\hat{a}_{\rm out}$ are the annihilation operators of the input and output optical fields, respectively, and $\gamma_{\rm opt}$ is the optomechanical anti-damping. The GW signal sideband phase delay $\Omega L_{\rm arm}/c$ can be compensated by the negative dispersion filter when $\gamma_{\rm opt} = c/L_{\rm arm}$, so long as we remain in the linear negative dispersion regime $\Omega \ll \gamma_{\rm opt}$. For a 4 km interferometer, $\gamma_{\rm opt}/(2\pi) = 12$ kHz. The pumping power necessary to achieve the desired $\gamma_{\rm opt}$ is determined by $\gamma_{\rm opt} = g^2/\gamma_f$, where $\gamma_f$ is the filter cavity bandwidth. The optomechanical coupling rate can be expanded into $g = \dfrac{d\omega}{d q} x_{\rm zpf}$ $\bar{a}$, where $\dfrac{d\omega}{d q}$ is the optical frequency shift per unit of generalised mechanical displacement $q$, $x_{\rm zpf}$ is the mechanical zero-point displacement fluctuation and $\bar{a}^2$ is the mean intracavity photon number. The power requirement for the negative dispersion filter becomes:

\begin{equation}\label{eq:antidampingPowerRequirement}
    P_f = \dfrac{ c \,M_{\rm eff} \,\omega_p \, \omega_m \gamma_{\rm opt}\gamma_f}{L_f}\dfrac{1}{\left(d\omega/dq\right)^2},
\end{equation}

\noindent
where $M_{\rm eff}$ is the effective mass of the mechanical resonator and $\omega_p = \omega_0 + \omega_m$ is the pump frequency of the filter. The optomechanical coupling is found from the relation of optical resonant frequency versus membrane displacement for a MIM cavity, given as \cite{JayichMIM, ThompsonMIM}:

\begin{equation}\label{eq:mimCoupling}
    \omega(x) = (c/L_f)\arccos{\left(\left|r_m\right|\cos{\left(4\pi x/\lambda\right)}\right)}
\end{equation}

\noindent
where $r_m$ is the membrane amplitude reflectivity and $x$ is the membrane displacement. The optical frequency $\omega(x)$ is periodic with $x$ in the MIM cavity. A 20 nm layer of silicon nitride with refractive index $n_m =2.0$ will have a power reflectivity of $r_m^2 = 0.03$ at $\lambda = 1064$ nm wavelength (see figure 6.1 in \cite{WilsonThesis}). This is low compared to dielectric-stack Bragg reflectors, but nevertheless is still enough to obtain sufficient optomechanical coupling.

\begin{figure}
    \centering
    \includegraphics[width=0.5\textwidth]{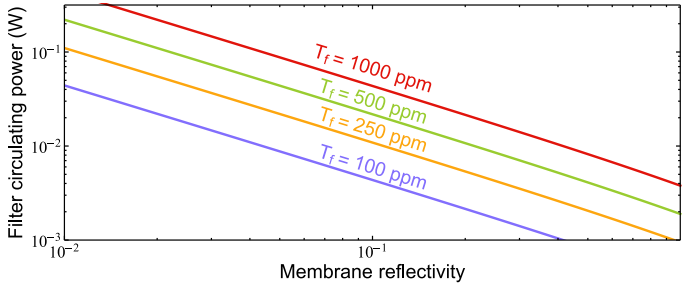}
    \caption{\textbf{Intracavity power for negative dispersion filtering using phononic crystal resonator:} We wish to obtain $\gamma_{\rm opt}/2\pi = 12$ kHz using a phononic crystal membrane in a membrane-in-the-middle configuration. The required intracavity power is plotted versus the membrane power reflectivity for different input coupler transmissivities $T_f$. }
    \label{fig:pncPower}
\end{figure}

We take the derivative of equation \ref{eq:mimCoupling} with respect to $q$, where $q$ in this case is equivalent to the membrane displacement $x$. Substituting into equation \ref{eq:antidampingPowerRequirement} gives the filter cavity power requirements shown in figure \ref{fig:pncPower}. We choose a filter cavity with input transmission $T_f = 300$ ppm to balance pumping power and quantum noise requirements, resulting in 42.2 mW circulating power. It has been shown that silicon nitride membrane resonators can maintain incident optical power approaching 0.1 W at around 10 K temperature \cite{MasonPNC}, so the calculated power requirement is plausible for the purpose of maintaining low $T/Q_m$. See Supplementary Material for more information regarding absorption heating.

Achieving WLSR using the BAW resonator characterised by Galliou \textit{et al.} requires coupling to the 204 MHz longitudinal mode in order to optimise frequency dependent $Q_m$ \cite{GalliouBAW}. However, the low optomechanical coupling rate of near-infrared light to this mechanical mode results in a required light intensity in excess of $10$ MW/cm$^2$. While quartz has been shown to withstand optical intensities greater than $1$ GW/cm$^2$ \cite{QuartzDT1,QuartzDT2}, maintaining high $Q_m$ at such high power has not been tested. In Supplementary Material we discuss optomechanical coupling of the BAW resonator and strategies that may be used to mitigate the extreme power requirement, such as alternate WLSR configurations and a single-layer quarter-wavelength coating. Also, the power requirement of equation $\ref{eq:antidampingPowerRequirement}$ scales inversely proportional to $L_{\rm arm}$, and will be reduced for interferometers such as Einstein Telescope and Cosmic Explorer that propose using arm lengths of 10 km or greater.

\section*{Noise spectrum of white light signal recycling interferometers}\label{sec:discussion}

\begin{figure}
    \centering
    \includegraphics[width=0.5\textwidth]{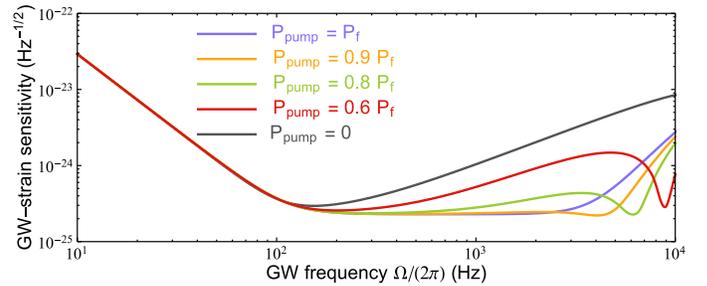}
    \caption{{\textbf{Demonstration of tunable quantum noise spectrum in WLSR interferometer:}} Adjustment of the quantum noise curve is achieved by changing the filter cavity pump power $P_{\rm pump}$ relative to $P_f$, which is the power required to achieve $\gamma_{\rm opt}=c/L$. This is useful for tuning the location of peak sensitivity in the neutron star detection band without changing interferometer hardware or detuning the signal recycling cavity.}
    \label{fig:tunableFilter}
\end{figure}

We calculate the noise spectrum of figure \ref{fig:layout} using the two photon quantum optics formalism of Caves and Schumaker \cite{CavesTwoPhoton1, CavesTwoPhoton2, CorbittTwoPhoton}. Cavity components are represented as transfer matrices which incorporate optomechanical interaction \cite{CorbittTwoPhoton, HaixingFilters}. The basis vector consists of the amplitude and phase quadratures of the light field. Transfer matrices of the cavity components are then multiplied to obtain an overall transfer function. We assume homodyne phase quadrature measurement of the interferometer output beam with no detuning of the signal recycling cavity.

The main noise input into the calculation shown in figure \ref{fig:layout} are as follows. Thermal noise from the mechanical resonator is introduced as displacement (phase) noise imparted onto the beam inside the filter cavity. Optical losses are input as uncorrelated vacuum in both the amplitude and phase quadratures. We introduce optical losses in the arm cavity, filter cavity, output train and beamsplitter cavity. Beamsplitter cavity losses are dependent upon the incident power on the interferometer beamsplitter, while other losses are assumed to be power-independent. Quantum noise from filter cavity sidebands at $\omega_0+2\omega_m\pm\Omega$ is also present, and its effect on the interferometer signal is suppressed using a high finesse filter cavity and high $\omega_m$. Further details are presented in Methods, and tables of values are given in Supplementary Material.

WLSR using PNC resonators is seen to reach strain sensitivity levels below $h\sim 10^{-24}$ Hz$^{-1/2}$ at GW signal frequencies up to 5 kHz, with a peak sensitivity below $6 \times 10^{-25}$ Hz$^{-1/2}$ across a broad band, as seen in figure \ref{fig:layout}. This particular GW detector configuration uses 4 MW arm cavity power, 10 dB frequency dependent squeezing and superior optical loss performance compared to the planned near-future upgrade of the current Advanced LIGO network known as A+ \cite{LIGOWP2019}. The peak sensitivity is limited by thermal noise coupling, set by the Q-factor $Q_m = 1.03 \times 10^9$ at temperature $T=1$ K. At frequencies of 1--5 kHz, we are also concerned with optical loss from the filter cavity. The PNC sensitivity curve of figure \ref{fig:layout} is set with 10 ppm filter cavity round trip loss as a desirable target. Previously reported measurements of silicon nitride absorption indicate that 1--4 ppm or lower absorption is possible in the case of a 20 nm thick resonator and 1064 nm wavelength light \cite{WilsonThesis, SankeyNonlinear}.

In longer interferometers, quantum shot noise scales with $1/\sqrt{L}$, whereas classical displacement noises scale with $1/L$. Breaking the sensitivity/bandwidth compromise using WLSR is a potential strategy to bring quantum noise down to the level of classical noises in future detectors that plan to use arms of 10 km length or greater. However, the filter cavity thermal noise requirement is proportional to the unmodified interferometer bandwidth, and is thus stricter for longer interferometers.
 
An interesting nuance in the quantum noise spectrum is present when the pumping power is not perfectly matched to $\gamma_{\rm opt}$, as shown in figure \ref{fig:tunableFilter}. The quantum noise spectrum exhibits a small region of enhanced high frequency sensitivity, at the expense of some lower frequency sensitivity. This insight provides an additional advantage for the detection of gravitational waves from binary neutron star coalescence, since the exact frequency of the kilohertz ringdown is unknown. WLSR presents the possibility of shifting the optimal detection frequency of the interferometer without changing the interferometer hardware or detuning the signal recycling cavity. However, this is contingent on maintaining a low contribution of filter cavity optical loss, which is easier for longer interferometers, as demonstrated in Supplementary Material.

\section*{Discussion}

There are several technical concerns not directly considered in the above calculations, but which will be important for implementing WLSR.

The parametric interaction between the two light fields and mechanical motion in the filter cavity results in optomechanical instability that must be controlled. Previous analysis has shown a state space demonstration of filter cavity controllability without imparting extra noise on the overall sensitivity \cite{HaixingFilters}. However, a later analysis showed that accounting for the time delay of the control system results in readout noise \cite{JoeTransmission}.

Mode mismatch between the interferometer and negative dispersion would introduce signal recycling cavity loss. While the noise budgets shown in figure \ref{fig:layout} account for general signal recycling loss, the specific contribution of mode matching and angular alignment control have yet to be investigated. However, it is expected that the contribution will be significant given the laser beam size in the filter. From the selection of experimentally demonstrated high-Q PNC resonators sufficient for WLSR, the largest is 350 $\mu$m wide \cite{DenmarkPNC}, requiring a beam waist of approximately 80-100 $\mu$m to reduce optical diffraction losses. For the BAW resonator, the effective width of the 204 MHz longitudinal mode is 260 $\mu$m. The beam waist in the Advanced LIGO output mode cleaner is $\sim$ 500 $\mu$m~\cite{Fricke_2012}, and Advanced LIGO target~1--2\% mode matching losses for the next generation of 4 km detectors \cite{LIGOWP2019}. The impact of this level of loss relative to quantum and thermal noise is shown in Supplementary Material.   
 
Scattered light rejoining the interferometer beam can contaminate the signal with phase noise acquired from moving objects. An estimate can be obtained by analysing the degree of freedom along the optical axis. The mechanical resonator motion must be controlled, so the dominant contribution is from the filter cavity vacuum enclosure. Assuming that the motion is typical of LIGO isolated tables, the maximum tolerable light power rejoining the interferometer beam is approximately 0.5 ppm of light power incident upon the filter cavity \cite{HiroScatterLight}. In figure \ref{fig:layout}, a large window to the cryogenic component is avoided for this reason.

Silicon nitride phononic crystal resonators provide a first realistic means of creating a white light signal recycling interferometer, using experimentally demonstrated values of mechanical loss, optical absorption and incident laser power. Single crystal bulk acoustic wave resonators also have promising thermal noise properties, but the required levels of optical power are untested. Proposed long-arm detectors such as Einstein Telescope and Cosmic Explorer will relax the optical power requirement, giving us more flexibility in future designs. An optomechanical negative dispersion filter for WLSR is currently under development at the University of Western Australia. The properties of silicon nitride phononic crystal resonators make them ideal for promptly achieving practical broadband enhancement of GW detector sensitivity, allowing greater investigation of neutron star coalescence.

  \section*{Methods}
  
  \subsection*{Calculation of sensitivity spectrum for WLSR interferometers}

Noise budgets of WLSR interferometers are calculated using the two photon formalism of Caves and Schumaker \cite{CavesTwoPhoton1, CavesTwoPhoton2, CorbittTwoPhoton}, where cavity components are represented as transfer matrices which can also incorporate optomechanical interaction \cite{CorbittTwoPhoton, HaixingFilters}. For example, a beam travelling distance $L_{\rm free}$ and reflected from a moving mirror in free space can be represented as:

\begin{equation}\label{eq:twoPhotonFreeSpace}
\begin{bmatrix}
\hat{\beta}_a (\Omega) \\
\hat{\beta}_p(\Omega)
\end{bmatrix} = e^{2\mathbf{i}\Omega L_{\rm free}/c}\begin{bmatrix}
1 & 0\\
-\kappa & 1
\end{bmatrix}\cdot\begin{bmatrix}
\hat{\alpha}_a (\Omega) \\
\hat{\alpha}_p(\Omega)
\end{bmatrix},
\end{equation}
\noindent
where $\alpha$ and $\beta$ respectively  represent the input and output beams, and subscripts $a$ and $p$ the amplitude and phase quadratures. The amplitude and phase quadratures of light are related to the sideband creation and annihilation operators by:

\begin{equation}\label{eq:sidebandToQuadrature}
    \begin{bmatrix}
\hat{\alpha}_a (\Omega) \\
\hat{\alpha}_p(\Omega)
\end{bmatrix} =  \dfrac{1}{\sqrt{2}}\begin{bmatrix}
1 & 1\\
\mathbf{i} & -\mathbf{i}
\end{bmatrix}  \begin{bmatrix}
\hat{\alpha}_-^{\dag} (\Omega) \\
\hat{\alpha}_+(\Omega)
\end{bmatrix},
 \end{equation}
 \noindent
where $\hat{\alpha}_+(\Omega)$ is the annihilation operator of the upper sideband at frequency $+\Omega$ with respect to the reference and $\hat{\alpha}_-^{\dag}(\Omega)$ is the creation operator of the lower sideband at frequency $-\Omega$ with respect to the reference. As such the two-photon formalism is naturally used in cases where modulation produces paired sidebands. Optomechanical coupling is incorporated in the frequency of mirror motion $\Omega$ and the coupling factor $\kappa = 8 P_0 \omega_0 / (M c L_{\rm free})$, where $P_0$ is the incident power and $M$ is the mass of the mirror. $2\times2$ transfer matrices in the two-photon basis can also be built up for tuned and detuned optomechanical cavities in a similar manner to equation \ref{eq:twoPhotonFreeSpace}. We obtain an overall sensitivity spectrum by looking at the input-output relation at the output photodetector. For this transfer matrix method it is simple to calculate the sensitivity spectrum for the measurement of any linear combination of amplitude and phase quadrature, though for the purpose of this paper we only require measurement of the phase quadrature.

\begin{figure}
    \centering
    \includegraphics[width=0.45\textwidth]{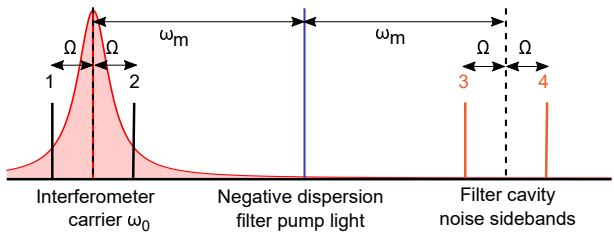}
    \caption{\textbf{Optical frequencies present in the negative dispersion filter:} The gravitational wave signal sidebands, labelled ``1'' and ``2'', are detuned from the interferometer carrier frequency $\omega_0$ by the GW signal frequency $\Omega$. The resonance peak of the interferometer and filter cavity is shown centered around $\omega_0$. The filter cavity is pumped by blue detuned light at frequency $\omega_0 + \omega_m$. Additional noise sidebands ``3'' and ``4'' are present at $\omega_0 + 2\omega_m \pm \Omega$, which couple to the GW signal sidebands ``1'' and ``2'', introducing extra quantum noise at $\omega_0 \pm \Omega$.}
    \label{fig:sidebandScheme}
\end{figure}

Additional noise sidebands are produced by the filter cavity, as illustrated by figure \ref{fig:sidebandScheme}. The calculation considers the GW signal sidebands at optical frequencies of $\omega_0 \pm \Omega$ along with the noise sidebands $\omega_0 + 2 \omega_m \pm \Omega$ that arise as a result of radiation pressure interactions of the GW sidebands within the detuned filter cavity. Doing so requires expanding the optomechanical transfer matrix from $2\times2$ to $4\times4$, and for the detuned filter cavity, we also switch from the two-photon picture to the sideband creation/annihilation picture. The basis vector incorporates each of the sidebands shown in figure $\ref{fig:sidebandScheme}$. As such, we produce a transfer function:
\begin{equation}
\mathbf{\beta_{4\times1}}(\omega) = \mathbf{M}_{4\times4}(\omega)\cdot\mathbf{\alpha_{4\times1}}(\omega),
\end{equation}
where $M_{\rm 4\times4}$ is the transfer matrix of an optomechanical cavity detuned from $\omega_0$ by $\omega_m$, and the argument $\omega$ denotes the separation of sidebands that appear centered around $\omega_0+\omega_m$, as per figure \ref{fig:sidebandScheme}. For example, sideband 1 in figure \ref{fig:sidebandScheme}, the lower GW signal sideband, is separated from the center frequency by $\omega = \omega_m+\Omega$. An appropriate transfer matrix can be constructed by taking the two-photon transfer matrix of an optomechanical cavity detuned by $\omega_m$, with GW sidebands occurring at $\omega = \omega_m\pm\Omega$, transforming to the sideband basis using the matrix in equation \ref{eq:sidebandToQuadrature}, and arranging the appropriate entries into a $4\times4$ matrix according to the following basis:

\begin{equation}
    \begin{bmatrix}
    \hat{\beta}_-^{\dag}(\omega_m+\Omega) \\
    \hat{\beta}_-^{\dag}(\omega_m-\Omega) \\
    \hat{\beta}_+(\omega_m-\Omega) \\
    \hat{\beta}_+(\omega_m+\Omega)
    \end{bmatrix} = \mathbf{M}_{4\times4}(\omega)\cdot  \begin{bmatrix}
    \hat{\alpha}_-^{\dag}(\omega_m+\Omega) \\
    \hat{\alpha}_-^{\dag}(\omega_m-\Omega) \\
    \hat{\alpha}_+(\omega_m-\Omega) \\
    \hat{\alpha}_+(\omega_m+\Omega)
    \end{bmatrix}.
\end{equation}

Conjugating the second and third rows of the transfer matrix $\mathbf{M}_{4\times4}$ represents changing the second entry of the basis vector to an annihilation operator and the third entry to a creation operator. This allows us to use the following transformation matrix:

\begin{equation}
\begin{bmatrix}
    \hat{\alpha}_a(\Omega) \\
    \hat{\alpha}_p(\Omega) \\
    \hat{\alpha}_a(2\omega_m-\Omega) \\
    \hat{\alpha}_p(2\omega_m+\Omega)
    \end{bmatrix}=\dfrac{1}{\sqrt{2}}
    \begin{bmatrix}
    1 & 1 & 0 & 0\\
    \mathbf{i} & \mathbf{-i} & 0 & 0 \\
    0 & 0& 1 & 1 \\
    0 & 0 & \mathbf{i} & \mathbf{-i}
        \end{bmatrix}
        \cdot
  \begin{bmatrix}
    \hat{\alpha}_-^{\dag}(\Omega) \\
    \hat{\alpha}_+(\Omega) \\
    \hat{\alpha}_-^{\dag}(2\omega_m-\Omega) \\
    \hat{\alpha}_+(2\omega_m+\Omega)
    \end{bmatrix},
    \end{equation}
\noindent
where the frequency of the argument is now written with respect to the carrier frequency $\omega_0$ instead of the blue-detuned pumping frequency $\omega_0+\omega_m$. The first two rows represent the amplitude and phase quadrature of signal sidebands generated about $\omega_0\pm \Omega$, while the third and fourth rows represent the quadratures of light generated about $\omega_0 + 2\omega_m \pm \Omega$. The transfer matrix $\mathbf{M}_{4\times4}$ can thus give the two-photon transfer function for the quantum noise from sidebands at $\omega_0 + 2\omega_m \pm \Omega$. For simplicity, we assume that the sidebands at $\omega_0+2\omega_m\pm\Omega$ are far enough detuned from the interferometer resonance to be simply reflected back into the signal recycling optics.

As a consequence of keeping propagation phase factor such as that shown in equation \ref{eq:twoPhotonFreeSpace}, the calculation also takes into effect cavity free spectral range, which has a significant impact on the audio band sensitivity of GW detectors 10 km and above in length. 

An initial impression dictates that $\gamma_{\rm opt}/(2\pi)$ is set to $12$ kHz in order to cancel the phase delay accumulated by GW signals in the 4 km interferometer arms. However, by slightly offsetting the optomechanical antidamping, the calculated quantum noise response curve extends further into the 1--5 kHz NS band at a slight cost in peak sensitivity, as indicated by figure \ref{fig:tunableFilter}. 

Optical losses from various sources introduce uncorrelated vacuum noise to the GW signal sidebands. Optical loss from the negative dispersion filter is treated as transmission of uncorrelated vacuum through the end mirror of the filter cavity. Likewise, optical loss in the interferometer arms is introduced as transmission of uncorrelated vacuum through the end test mass. Loss from the output optics to the photodiode is introduced between the SRM and output Faraday isolator. It behaves similar in frequency dependence to the quantum noise curve, but is actually caused by the homodyne detection process as described by Kimble \textit{et al.} \cite{KimbleSqueezing}. Similar to Martynov, \textit{et al.}, we consider the effect of resonantly enhanced optical losses inside the interferometer beamsplitter cavity, which is dominated by power-dependent thermal lensing noise \cite{DenisHF}. This is due to absorption of optical power onto the ITM and beamsplitter, causing heat gradients that distort the carrier wave from its desired shape. These losses are then resonantly enhanced inside the beamsplitter cavity. In the WLSR configuration of figure \ref{fig:layout}, this resonant enhancement of arm power-dependent optical loss occurs inside the SEM/ITM cavity. The wavefront distortion contributions from the ITM and beam splitter scale approximately as \cite{DenisHF, BrooksAdaptive}:

\begin{align}
    \epsilon_{\rm ITM} &= \left(\dfrac{P}{1 \mathrm{ MW}}\dfrac{\alpha_{\rm ITM}}{0.5 \mathrm{  ppm}}\dfrac{30}{\kappa_{\rm ITM}}\right)^2 \times 1000 \mathrm{ ppm} \\
    \epsilon_{\rm BS} &= \left(\dfrac{P_{\rm BS}}{6 \mathrm{ kW}}\dfrac{\alpha_{\rm BS}}{1 \mathrm{ ppm}}\dfrac{1}{\kappa_{\rm BS}}\right)^2 \times 250 \mathrm{ ppm}, 
\end{align}

\noindent
where $\alpha_{\rm ITM, BS}$ represent optical absorption, $\kappa_{\rm ITM, BS}$ the compensation factor from various systems that reduce thermal lensing and $P_{\rm BS}$ the incident power on the beamsplitter. The total signal extraction loss $\epsilon_{\rm se} = \epsilon_{\rm ITM} + (\epsilon_{\rm BS}/2)$ is introduced as uncorrelated vacuum between the main beamsplitter and SEM. Resonant enhancement causes significant contribution of signal extraction loss in the 1--5 kHz band.

Introducing the SEM also causes impedance matching of losses between the signal recycling cavity and the arm cavity. As such, losses occurring in the signal recycling cavity (SRC) are combined with arm cavity losses into one total loss $\epsilon_{\rm arm}$, which is introduced as uncorrelated vacuum inside the SRC. The shot noise power spectrum of this loss behaves similarly to that of quantum noise in a simple Michelson at high frequency, scaling inversely proportional to $L_{\rm arm}^2$. By contrast, the shot noise power spectrum of resonantly enhanced optical losses scales inversely proportional to $L_{\rm arm}$. As such, the relative contribution of impedance matched arm losses is reduced in longer interferometers.

Specific values used for the WLSR interferometer sensitivity calculations are tabulated in Supplementary Material.

  \section*{Code availability}
  Calculations regarding the noise budget of WLSR interferometers were performed using Mathematica. Annotated code is available from the corresponding author upon request.
  
 \section*{Author contributions}

Calculations and models regarding GW detector interferometry and filter cavity optomechanics were performed by MAP, and discussed and verified by HM, YM, YC and CZ. CDB and DGB provided discussion on the integration of WLSR in GW detectors. DM, MR and AS provided information regarding measured data of PNC resonators. MG and MET provided information regarding measured data of BAW resonators. CZ, LJ and DGB were the main supervisors of the project. The paper was drafted by MAP, MG, YC, AS and DGB, edited by MAP and commented by all authors.
 
 \section*{Acknowledgements}
 This research was primarily supported by the Australian Research Council Center of Excellence for Gravitational Wave Discovery OzGrav CE170100004 and Discovery Project DP170104424. In addition, CDB is supported by the Discovery Early Career Researcher Award DE190100437. The work of DM, MR and AS was supported by the European Research Council project Q-CEOM (grant no. 638765) and the EU H2020 FET proactive project HOT (grant no. 732894). MG and MET are supported by the ARC Center of Excellence for Engineered Quantum Systems EQuS CoE  CE170100009. YC is supported by the US NSF Grants PHY-1708212 and PHY-1708213, and by the Simons Foundation (Award Number 568762). HM is supported by UK STFC Ernest Rutherford Fellowship (Grant No. ST/M005844/11).

\bibliography{bulkRefs}

\end{document}

\title{Supplementary Material for ``Gravitational wave detectors with broadband high frequency sensitivity"}

\author{Michael A. Page\affmark[1], Maxim Goryachev\affmark[2], Haixing Miao\affmark[3], Yanbei Chen\affmark[4], Yiqiu Ma\affmark[5], David Mason\affmark[6], Massimiliano Rossi\affmark[7], Carl D. Blair\affmark[1], Li Ju\affmark[1], David G. Blair\affmark[1], Albert Schliesser\affmark[7], Michael E. Tobar\affmark[2], and Chunnong Zhao\affmark[1]}

\affiliation{\affmark[1]Australian Research Council Center of Excellence for Gravitational Wave Discovery, University of Western Australia, 35 Stirling Highway, Perth, Western Australia 6009, Australia}
\affiliation{\affmark[2] Australian Research Council Center of Excellence for Engineered Quantum Systems, University of Western Australia, 35 Stirling Highway, Perth, Western Australia 6009, Australia}
\affiliation{\affmark[3] Astrophysics and Space Research Group, University of Birmingham, Birmingham B15 2TT, United Kingdom}
\affiliation{\affmark[4] Theoretical Astrophysics, California Institute of Technology, 1200 E California Blvd, Pasadena, California 91125, United States}
\affiliation{\affmark[5] Center for Gravitational Experiment,  School of Physics, Huazhong University of Science and Technology, Wuhan, 430074, China}
\affiliation{\affmark[6] Yale Quantum Institute, Yale University, 17 Hillhouse Ave, New Haven, Connecticut 06511, United States
}
\affiliation{\affmark[7] Niels Bohr Institute, University of Copenhagen, Blegdamsvej 17, 2100 Copenhagen, Denmark}

\maketitle

\section*{Optomechanical white light signal recycling requirements}\label{sec:wlsr}

Here we go further into detail on the theoretical background of the negative dispersion filter, outlined in the section ``Optomechanical coupling in negative dispersion filters'' of the main text.

The optomechanics of the negative dispersion filter can be characterised by a Hamiltonian given by:

\begin{equation}
    \hat{H} = \hat{H}_{\rm opt} + \hat{H}_{\rm mech} + \hat{H}_{\rm int} + \hat{H}_{\rm ext}^{\rm opt} + \hat{H}_{\rm ext}^{\rm mech}.
    \end{equation}
    \noindent
The interaction component is given by the following form:

\begin{equation}\label{eq:Hint}
    \hat{H}_{\rm int} = -\hbar \dfrac{d \omega}{d q} x_{\rm zpf}(\hat{b}+ \hat{b}^{\dag})\hat{a}^{\dag}\hat{a},
\end{equation}
\noindent
The free Hamiltonians for the optical and mechanical resonances inside the filter cavity are denoted by $\hat{H}_{\rm opt}$ and $\hat{H}_{\rm mech}$. Langevin coupling to the external bath is denoted by $\hat{H}_{\rm ext}^{\rm opt}$ and $\hat{H}_{\rm ext}^{\rm mech}$ \cite{GardinerLangevin, YanbeiReview}. The filter cavity operates by parametric interaction with gravitational wave signal fields at frequency $\omega_0 \pm \Omega$. To mediate this interaction, the filter pump is blue detuned from the interferometer carrier by $\omega_m$. In this scheme, the optomechanical coupling Hamiltonian can be expressed in a linearised form using the rotating wave approximation in the interaction picture, resulting in equation 1 of the main text.

After applying Heisenberg's equations of motion, transferring the differential equations to the frequency domain and removing small terms, the GW signal transfer function is found. GW sidebands of frequency $\Omega$ acquire a frequency dependent negative phase that depends on the optomechanical anti-damping $\gamma_{\rm opt}$. By matching $g$ to $\gamma_{\rm opt}$ and filter cavity bandwidth $\gamma_f$ such that $\gamma_{\rm opt} = g^2/\gamma_f$, the GW signal sideband transfer becomes \cite{HaixingFilters}:

\begin{equation}\label{eq:negativeDispersion}
    \hat{a}_{\rm out}(\omega_0 +\Omega) = \dfrac{\Omega+i\gamma_{\rm opt}}{\Omega-i\gamma_{\rm opt}}\hat{a}_{\rm in}(\omega_0 +\Omega) = e^{-2i\phi}\hat{a}_{\rm in}(\omega_0 +\Omega),
\end{equation}

\noindent where $\phi = \arctan\left(\Omega/\gamma_{\rm opt}\right)$. The optomechanical coupling can be tuned such that the negative dispersion phase $\phi$ matches the GW signal sideband phase delay $\Omega L_{\rm arm}/c$ acquired from travel inside the interferometer arm cavities, which implies $\gamma_{\rm opt} = c/L_{\rm arm}$. For example, a 4 km long GW detector requires $\gamma_{\rm opt}/(2\pi) = 12$ kHz. The linear negative dispersion regime  $\phi \sim \Omega/\gamma_{\rm opt}$ applies for $\Omega \ll \gamma_{\rm opt}$, where the transfer function can be simplified to equation 2 shown in the main text. 

Extra vacuum noise sidebands are present at $\omega_0 + 2 \omega_m \pm \Omega$ as shown in figure 4 of the main text Methods. These interact to produce extra noise at the gravitational wave sideband frequency. Solving for the transfer function of a detuned cavity, it is possible to show that the noise sidebands around $\omega_0 +2\omega_m$ have a first order contribution proportional to $\gamma_f/\omega_m$, such that:

\begin{align}\label{eq:ndNoise}
    \hat{a}_{\rm out}(\omega_0 +\Omega) = &e^{-2i\phi}\hat{a}_{\rm in}(\omega_0 +\Omega) \nonumber \\&- \dfrac{\gamma_f}{\omega_m}\dfrac{\gamma_{\rm opt}}{\Omega-i\gamma_{\rm opt}}\hat{a}(\omega_0 +2\omega_m-\Omega),
\end{align}

\noindent
where in this case the sideband at frequency $\omega_0 +2\omega_m-\Omega$ contributes noise at frequency $\omega_0 +\Omega$. The extra noise sidebands of the detuned cavity introduce imperfect phase cancellation, reducing potential bandwidth enhancement from WLSR. Previous schemes for WLSR interferometry in GW detectors used pendulum resonators with $\omega_m/(2\pi) < 200$ kHz \cite{HaixingHF, PageWLSR}. The low mechanical frequency places strict requirements on the filter cavity finesse to reduce coupling of $\omega_0 + 2 \omega_m \pm \Omega$ noise sidebands to the GW signal. However, the high finesse filter cavity also increases the fractional contribution of filter cavity round trip optical loss $\epsilon_f$. The higher mechanical frequency of the resonators described in this paper allows significant suppression of sideband noise even for high bandwidth filters, which makes the optical cavity design requirements much more flexible.

The use of an optomechanical resonator in the filter cavity introduces thermal noise into the main interferometer, which must be minimised in order to maintain integrity of the sensitivity. The expectation value of the thermal bath operator is $\langle \hat{b}_{\rm th}^{\dag}\hat{b}_{\rm th}\rangle \sim (k_B T)/\hbar\omega_m$ inside the filter. The temperature and quality factor requirement can be derived from the equations of motion for the mechanical mode in the filter, \cite{HaixingFilters}:

\begin{equation}\label{eq:thermalNoise}
  \dfrac{T}{Q_m} < \dfrac{\hbar \gamma_{\rm ifo}}{8 k_B},  
\end{equation}
\noindent
where $\gamma_{\rm ifo}$ is a characteristic bandwidth that depends on the detector configuration. For example, in an Advanced LIGO type interferometer we must use a PNC resonator with a Q-factor of $1.03\times10^{9}$ operating near 1 K temperature.

 \section*{Mechanical loss of bulk acoustic wave resonators}

\begin{table}[]
    \centering
    \begin{tabular}{|c c c|}\hline
     Parameter    & Symbol & Value \\ \hline
       Refractive index (quartz)  & $n_Q$ & 1.54 \\
          Photoelastic constant (quartz)   & $p_{\rm 13}$ & 0.27 \\
         Density (quartz)  & $\rho$    & 2648 kg/m$^3$ \\
          Speed of sound (quartz) & $v_a$&6327 m/s\\
                    Crystal thickness & $q_0$ & 1 mm \\
                  Crystal radius & $r_c$ & 15 mm \\
                  Radius of curvature of convex face & $R$ & 300 mm \\
                 Longitudinal mode number &$m$ & 65 \\
                 Mechanical frequency  &  $\omega_m/(2\pi)$  & $204$ MHz \\
                 Mechanical Q-factor  &  $Q_m$  & 8$\times$10$^{9}$  \\ \hline
                \end{tabular}
    \caption{Material properties and dimensions of the BAW resonator characterised by Galliou, \textit{et al.} \cite{GalliouBAW}}
    \label{tab:crystalMode}
\end{table}

 In this section we detail the mechanical loss of BAW resonators at low temperature. In the event that the issues with low optomechanical coupling can be overcome, their low thermal noise is promising for broadband GW detectors.

A BAW resonator may be regarded as the phononic analogue of an optical Fabry-P{\'e}rot cavity. Typically they are made of a thin plate of a dielectric material which supports phonons of shear and longitudinal polarisations. Acoustic waves are reflected by the interface between vacuum and crystal, thus the thickness of the crystal sets the resonance conditions for different overtone modes.

BAW resonators have been shown to achieve extremely high $Q_m > 10^9$ at hundreds of MHz and cryogenic temperatures. For long-lived phonons at low temperature, internal mechanical loss of BAW resonators is dominated by crystal lattice anharmonicity in the Landau-Rumer regime where the mechanical quality factor is inversely proportional to temperature and independent of frequency \cite{landaurumer1}. The power law scaling of $Q_m$ with temperature has been experimentally demonstrated to be $Q_m~\sim~T^{-6.5}$ for quartz BAW resonators at liquid helium temperatures of 3--20 K \cite{Goryachev2013,GalliouBAW}. 

The optimal longitudinal mode number is dictated by the compromise between clamping loss and surface scattering loss. Energy leakage through the resonator support is a primary source of mechanical loss at low mode number. To overcome this effect in BAW devices, the crystal is designed with a plano-convex lens shape. The radius of curvature of the convex face creates a potential well that traps phonons in the central part of the disk. The corresponding distribution of acoustic energy has a Gaussian-like profile, thus reducing the energy loss into the support \cite{Stevens:1986aa}. Higher overtone modes typically result in better phonon trapping, since the mode amplitude at the edge of the crystal is smaller \cite{GoryachevGeometry}. However, at high overtone numbers frequency dependent scattering loss becomes dominant. Surface roughness scattering comes from imperfections on the surface layer of the crystal. The loss contribution increases as the acoustic wavelength approaches the imperfection size. Surface scattering has been found to scale as $Q_{\rm scattering} = \dfrac{2}{m}\cdot 10^{12}$ for quartz resonators. The optimal mode to balance support and scattering losses is found to have $m=65$, $\omega_m/( 2\pi) =204$ MHz, with $Q_m = 8\times 10^9$ at 4 K. The crystal used to produce this mode has properties listed in table \ref{tab:crystalMode}. To achieve improved Q-factor at higher mode number, one would need to suppress the standard deviation of the the surface roughness to a level better than 1-4 nm.

\begin{figure}
    \centering
    \includegraphics[width=0.45\textwidth]{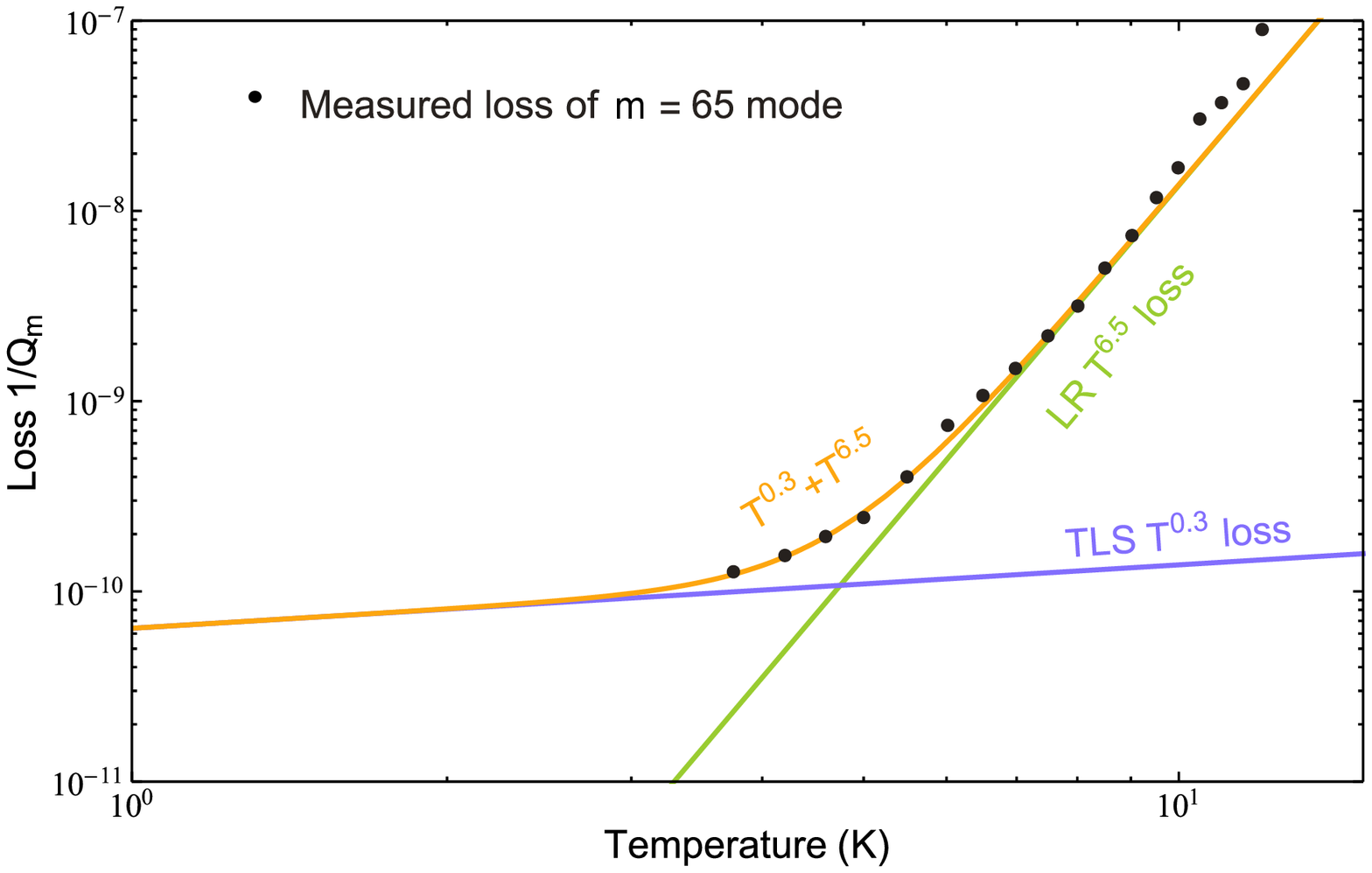}
    \caption{\textbf{Mechanical loss of quartz BAW resonator at cryogenic temperature:} Measured loss of the $m=65$ longitudinal mode of a quartz BAW resonator with TLS and Landau-Rumer (LR) losses. Between 5 and 10 K, the loss scales with $T^{6.5}$. At lower temperature, TLS loss for quartz is expected to follow a $T^{0.3}$ trend. Extrapolation of the temperature to 1 K indicates that $T/Q_m = 6.5 \times 10^{-11}$ K may be possible.}
    \label{fig:TLSQ}
\end{figure}

At temperatures of less than 1 K, the dominant temperature-scaling loss switches from the Landau-Rumer regime to intrinsic Two Level System (TLS) loss. Premium grade quartz crystals typically contain impurity ions such as Al$^{3+}$, Na$^+$, Si$^{4+}$, etc. at a concentration of a few parts per billion. TLS-limited Q-factor dependence on temperature is typically $Q_{\rm TLS}\sim T^{-0.3}$ for quartz crystals \cite{Goryachev2013,GalliouBAW}, which is supported by additional observations such as power dependence of losses, strong non-Duffing nonlinearities \cite{quartzJAP} and magnetic field sensitivity \cite{magneticBAW}. Comparison of experimental results with TLS theory gives a projected quality factor of the optimal BAW resonator mode shown in figure \ref{fig:TLSQ} \cite{GalliouBAW}. Extrapolating the TLS limited Q-factor reveals a limit of $Q_m = 1.6\times 10^{10}$ at 1 K temperature, which is used to produce the curve ``BAW Ultra-low loss'' in figure 1 of the main text.

Most sources of frequency noise for BAW resonators are significantly reduced in a typical cryogenic environment, leaving only the temperature fluctuations and external vibration as the prevailing factors \cite{instabilities2,instabilities}. Reduction of vibration sensitivity requires a stress compensated cut and crystal orientation with respect to main mode of vibration of the cryocooler \cite{instabilities}. Due to high degree of isolation of the acoustic wave from the environment and high operating frequencies, vibration from external sources has not been observed on any thermal noise spectra. 

 \section*{Negative dispersion optomechanics with bulk acoustic wave resonators}

\begin{figure}
    \centering
    \includegraphics[width=0.45\textwidth]{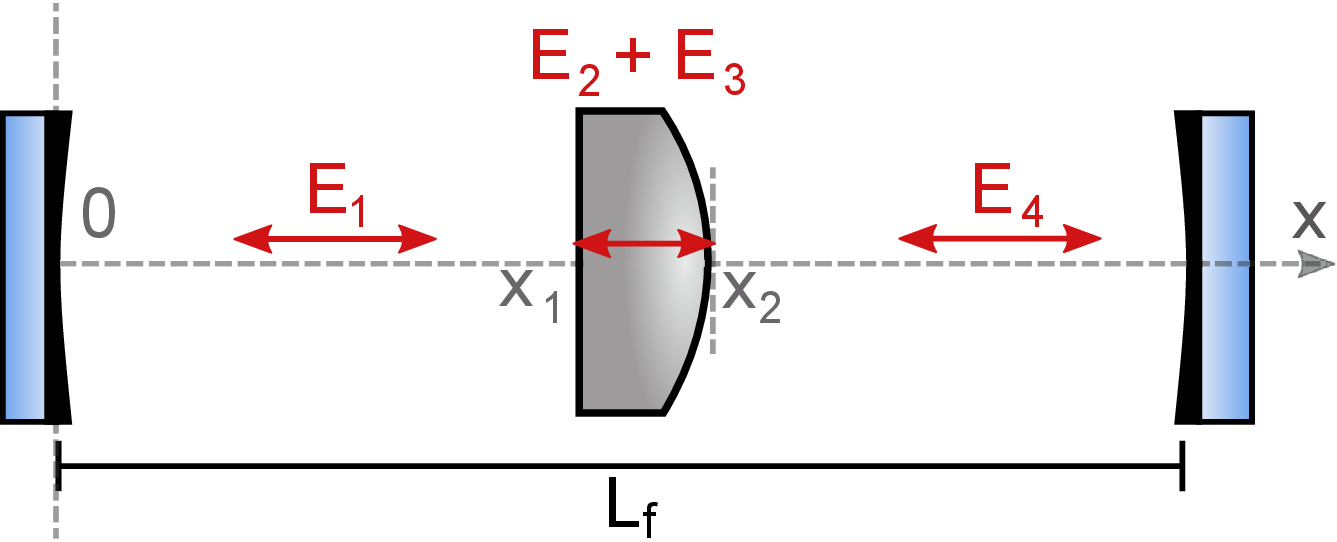}
    \caption{\textbf{Coupled optomechanical cavity containing a BAW resonator:} Illustration of the cavity parameters used in the calculation of optomechanical coupling to BAW resonator antiphase surface motion. The BAW resonator position is denoted using $x_1$ at the flat face and $x_2$ at the apex of the curved face. The electric field amplitude in the vacuum spaces is denoted by $E_1$ and $E_4$. The electric field inside the crystal is composed of a superposition of $E_2$ and $E_3$}
    \label{fig:BAWcoupling}
\end{figure}

In the main text, we demonstrated the possibility of creating WLSR using a PNC membrane resonator that interacts with light using the established method of membrane-in-the-middle optomechanics. The possibility of using millimeter-scale BAW resonators was also mentioned due to their appealing thermal noise properties, but there are outstanding issues with achieving the desired level of optomechanical coupling while maintaining a sufficient mechanical quality factor. Here we elaborate on the background and issues with optomechanical coupling of BAW resonators, in the context of WLSR GW detectors.

Optomechanical coupling of BAW resonators has been demonstrated using three-mode Brillouin scattering of optical waves to coherent acoustic phonons \cite{KharelBAW}. In order to achieve three-mode Brillouin scattering in the BAW resonator, two energy transfer conditions must be met. First, the optical frequencies must be separated by the mechanical frequency, which is achieved through design of the filter cavity. Second, the phase matching requirement states that $q_m = k_{j^{\prime}}+k_j$, where $q_m$ is the mechanical wavenumber and $k_{j,j^{\prime}}$ are the optical wavenumbers of the GW signal and blue detuned pump. For a three mode system in a quartz crystal, phase matching is satisfied when $\omega_m = 2 \omega_0 n_ Q v_a /c$, where $\omega_0$ is the frequency of 1064 nm wavelength light, $n_Q$ is the refractive index of quartz and $v_a$ is the speed of sound in quartz \cite{KharelBAW}. Using the properties shown in table \ref{tab:crystalMode}, the resulting mechanical frequency requirement of $\omega_m/(2\pi)\sim 18$ GHz is a factor of 89 higher than the 204 MHz mode with optimal $Q_m$. Since we wish to formulate a means of negative dispersion using the experimentally demonstrated Q-factor and frequency shown by Galliou \textit{et al.} \cite{GalliouBAW}, we must resort to another mechanism of optomechanical coupling.

In our calculations, the negative dispersion filter couples the optical modes of the cavity to the antiphase surface motion of the planar and convex faces of the BAW resonator crystal. This allows access to the highest measured Q-factor mode at 204 MHz. To model coupling of the light field to the surface vibrations of the BAW resonator, we consider a BAW resonator crystal situated inside a Fabry-Perot cavity, as shown in figure \ref{fig:BAWcoupling}. Using the boundary conditions of electromagnetic fields at the cavity end mirrors and crystal surfaces we obtain equations of the electric field as functions of crystal surface position. For given values of the cavity length $L_f$ and crystal center position $x_p$, we can obtain the cavity resonance frequency as a function of generalised displacement $q$, which in this case is the crystal thickness. Using electric and magnetic boundary conditions at the interface points $x=0, x_1, x_2, L_f$, we then construct a system of equations in terms of the electric fields $E_{\rm 1, 2, 3, 4}$, quartz refractive index $n_Q$, optical wavenumber $k$ and cavity length $L_f$.

\begin{align}\label{eq:BAWsystem}
&E_1 \sin(k x_1)=E_2\sin(n_Q k x_1)+E_3\cos(n_Q k x_1) \nonumber \\
&E_1 \cos(k x_1)=n_Q E_2\cos(n_Q k x_1)-n E_3\sin(n_Q k x_1) \nonumber \\
&E_4 \sin[k(x_2-L_f)]=E_2\sin(n_Q k x_2)+E_3\cos(n_Q k x_2) \nonumber \\
&E_4 \cos[k(x_2-L_f)]=n_Q E_2\cos(n_Q k x_2)-n_Q E_3\sin(n_Q k x_2). \nonumber \\
\end{align}
\noindent

The coordinates $x_1$ and $x_2$ are rearranged into crystal center position $x_p = (x_2 + x_1)/2$ and crystal thickness $q = x_2 - x_1$, and the system of equations reduces to:

\begin{align}
&\frac{n_Q\tan[k(x_p-q/2)]-\tan[n_Q k(x_p-q/2)]} {1+n_Q \tan[k(x_p-q/2)]\tan[n_Q k(x_p-q/2)]} \nonumber \\
=&\frac{n_Q \tan[k(x_p+\dfrac{q}{2}-L_f)]-\tan[n_Q k(x_p+\dfrac{q}{2})]} {1+n_Q \tan[k(x_p+\dfrac{q}{2})]\tan[k(x_p+\dfrac{q}{2}-L_f)]}.
\end{align}

\begin{figure}
    \centering
    \includegraphics[width=0.45\textwidth]{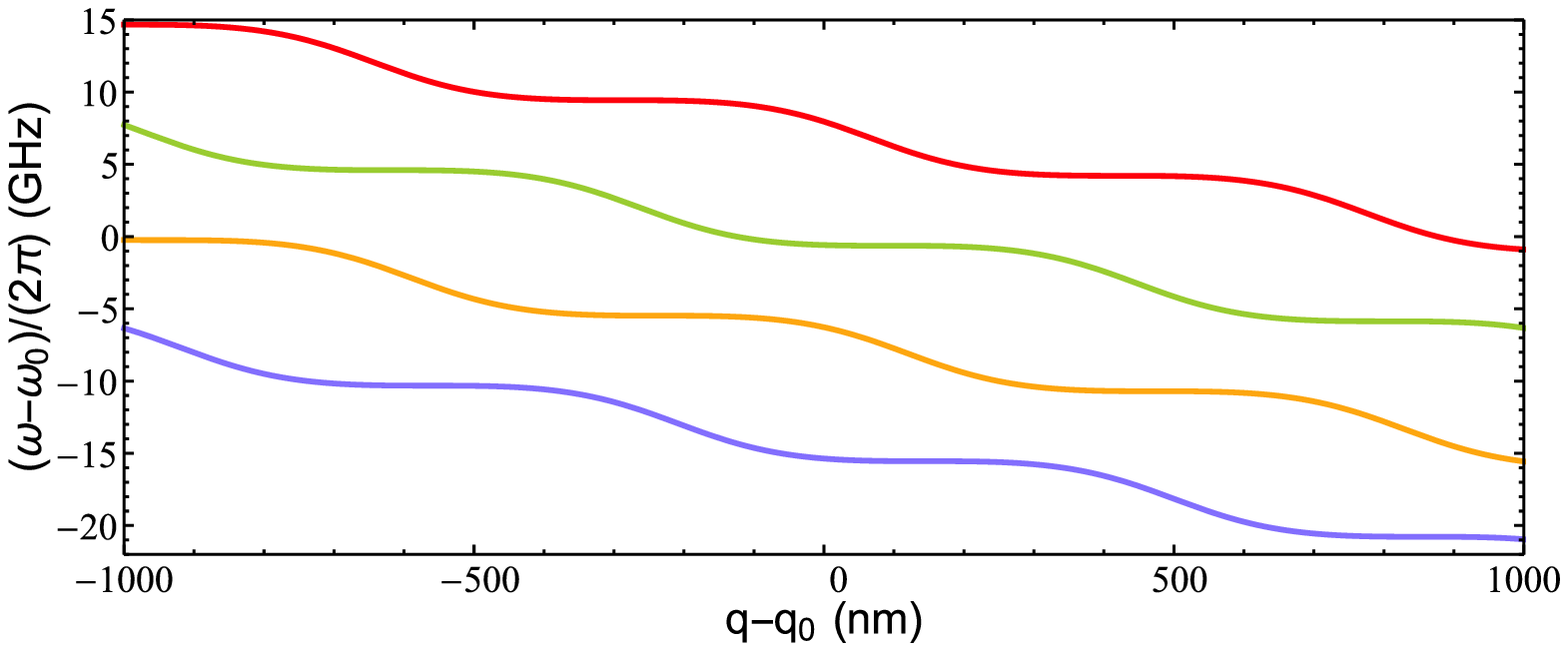}
    \caption{\textbf{Optical resonance for BAW resonator inside Fabry-Perot cavity:} Optical frequency $\omega$ versus thickness $q$ of the BAW resonator. The curves represent four selected optical modes, which are separated by the sinusoidally varying free spectral range of the cavity.}
    \label{fig:omc}
\end{figure}

\begin{figure}
    \centering
    \includegraphics[width=0.45\textwidth]{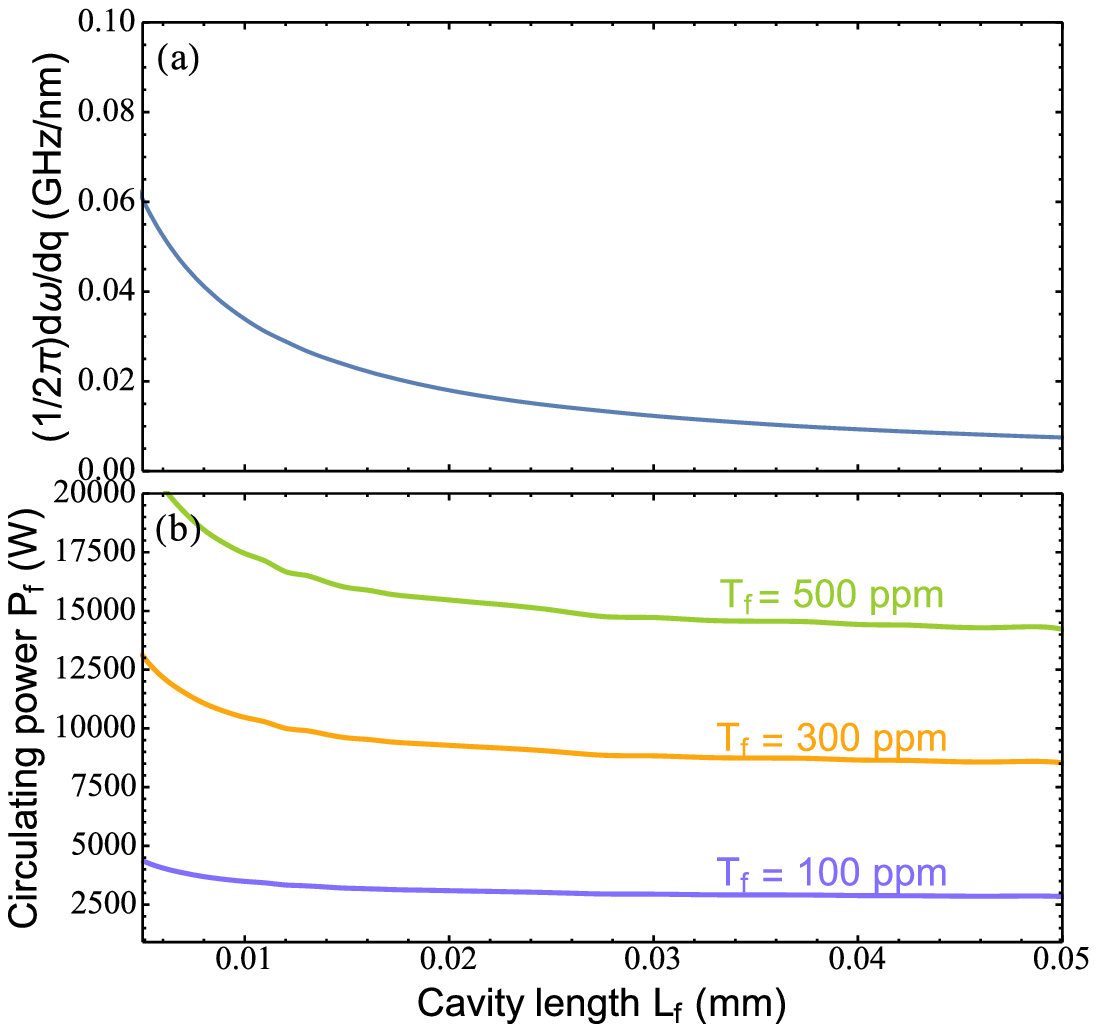}
    \caption{\textbf{Optomechanical coupling requirement for WLSR with BAW resonator:} The curves are calculated using a BAW resonator with initial thickness $q_0 = 1$ mm and center position $x_p = L_f/2$. (a) - Dependence of maximum optomechanical frequency shift $d\omega/dq$ on cavity length. (b) - Circulating power required to achieve $\gamma_{\rm opt}/(2\pi)=900$ Hz versus cavity length, for different values of filter input transmission $T_f$. This prescription is used for a WLSR configuration that produces somewhat lower bandwidth enhancement but also requires a factor of 13 less power versus the optimal WLSR described by equation 3 of the main text.}
    \label{fig:lengthPower}
\end{figure}

We then solve the wavenumber in terms of the generalised displacement $q$. Selected solutions are obtained in proximity to $\omega_0/(2\pi) = 2.82\times 10^{14}$ Hz, corresponding 1064 nm wavelength. The dependence of optical resonance frequency with crystal thickness is shown in figure \ref{fig:omc}. Over micron-scale motion, there is an approximately linear negative $d\omega/dq$, and an appropriate optical mode can be selected such that there is a linear negative $d\omega/dq$ within $-50$ nm $< q-q_0 < 50$ nm. Local frequency variation is due to the sloshing between the left coupled cavity, the crystal itself and the right coupled cavity. The free spectral range between modes varies sinusoidally, which is consistent with studies on BAW resonator coupled cavities \cite{KharelBAW}. The coupling $d\omega/dq$ decreases with cavity length as shown in figure \ref{fig:lengthPower}. For a small cavity of $L_f = 5$ mm, the maximum coupling is $d\omega/dq = 2\pi\times 0.061$ GHz/nm, while a longer cavity $L_f = 20$ mm results in maximum $d\omega/dq = 2\pi\times 0.018$ GHz/nm. The maximum single photon coupling rates $\dfrac{d\omega}{dq}x_{\rm zpf}$ are 0.10 Hz and 0.031 Hz, respectively. As expected, surface optomechanical coupling is small compared to optomechanical coupling to the bulk longitudinal mechanical mode, where Kharel \textit{et al.} demonstrated a near-infrared single photon coupling rate of 24 Hz in a BAW resonator at similar effective mass \cite{KharelBAW}. Micro-pendulums used in previous proposals \cite{PageWLSR} have estimated single photon coupling rates of $\sim$ 40 Hz, though at much lower effective mass ($\sim$ 10 ng) and mechanical frequency ($\sim$ 100 kHz).

To obtain the optomechanical antidamping necessary to produce negative dispersion, we must find the effective mass of the relevant mechanical mode. We use the following formula given by Goryachev \cite{GoryachevGeometry}:

\begin{equation}
    M_{\rm m, 0, 0} = \rho \pi \dfrac{q_0}{2}r_c^2 \dfrac{\mathrm{Erf}\left(\sqrt{m}\eta_x\right)\mathrm{Erf}\left(\sqrt{m}\eta_y\right)}{\eta_x \eta_y m},
\end{equation}
\noindent
where the crystal radius $r_c$ and density $\rho$ are given in table \ref{tab:crystalMode}. The coupling factors $\eta_{x, y}$ quantify the trapping of the Gaussian longitudinal mode within the crystal and are given by:

\begin{align}
   &\eta_x = r_c\sqrt{\pi \alpha} \\
   &\eta_y = r_c\sqrt{\pi \beta} \\
  & \alpha^2 = \dfrac{c_z}{R q_0^3 \mathcal{M}} \\
   &\beta^2 = \dfrac{c_z}{R q_0^3 \mathcal{P}},
\end{align}

\noindent
where $R$ is the radius of curvature of the curved surface of the BAW. $\mathcal{M}$ and $\mathcal{P}$ are material dependent transverse elastic parameters which are only well known at room temperature. Goryachev estimates $c_z/\mathcal{M} \sim c_z/\mathcal{P} \sim 0.4$ for cryogenic quartz crystals \cite{GoryachevGeometry}. This results in $\eta_x \sim \eta_y \sim 5.08$. The effective mass for the $m = 65$, $\omega_m/(2\pi) = 204$ MHz mode is $M_{\rm 65, 0, 0} = 0.56$ mg. In addition, the optical plane wave corresponding to this effective mass has area $A  = \dfrac{M_{\rm eff}}{\rho q_0}$ giving an appropriate optical beam radius of 260 $\mu$m to match to the mechanical mode.

\begin{figure}
    \centering
    \includegraphics[width=0.45\textwidth]{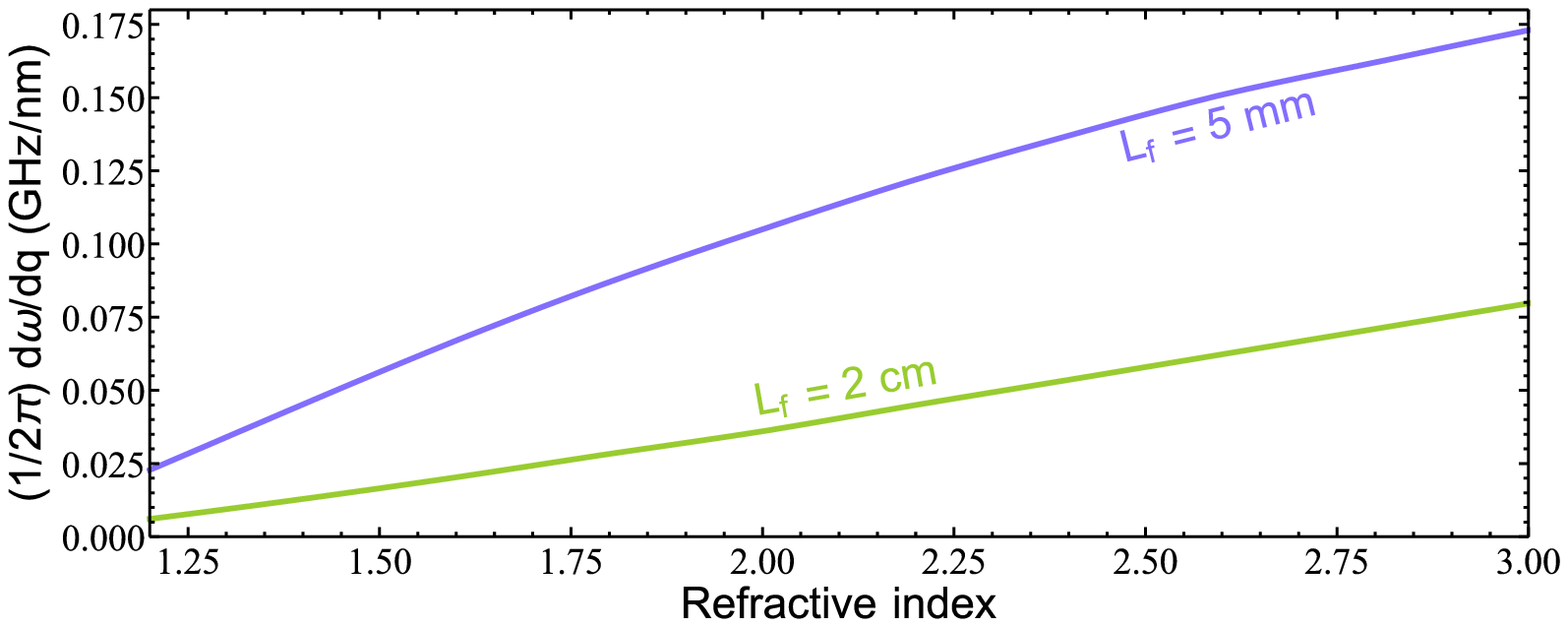}
    \caption{\textbf{BAW resonator optomechanical coupling versus refractive index}: Optomechanical frequency shift $d\omega/d q$ versus BAW resonator crystal refractive index, calculated using the system described in equation \ref{eq:BAWsystem} and figure \ref{fig:BAWcoupling}. Two different filter cavity lengths are shown - optomechanical coupling is smaller for longer cavities.}
    \label{fig:indexCoupling}
\end{figure}

Optomechanical coupling to the antiphase surface motion of a quartz resonator is relatively small. Even with a high finesse cavity of input transmission 100 ppm, for a cavity length of 50 mm, circulating power of 38 kW is required. This extreme power requirement is part of the motivation for exploring WLSR using less $\gamma_{\rm opt}/(2\pi)\sim  1000$ Hz in exchange for a slightly reduced bandwidth of sensitivity enhancement. The relationship of $P_f$ versus $L_f$ is shown in figure \ref{fig:lengthPower}, using $\gamma_{\rm opt}/(2\pi) = 900$ Hz specifically chosen for this alternate filter design. It is seen that $L_f < 10$ mm greatly increases the requirement on $P_f$. Decreasing $T_f$ is seen to decrease the required filter cavity power, and as per equation \ref{eq:ndNoise} also decreases the contribution of quantum noise from $\omega_0 + 2\omega_m \pm \Omega$ sidebands. However, the fundamental bandwidth broadening effect of WLSR scales with with $\gamma_f$ \cite{JoeTransmission}, and $T_f$ in the range 100--1000 ppm is chosen to balance these factors. Choosing $T_f = 100$ ppm gives a power requirement of approximately 2.5 kW, resulting in a beam intensity of 12 MW/cm$^2$. The damage threshold of quartz with respect to near-infrared light is reportedly greater than $1$ GW/cm$^2$ \cite{QuartzDT1, QuartzDT2}. However, it remains to be seen whether or not the power and intensity levels can be sustained in a cryogenic high-Q resonator. Possible issues may include wavefront distortion and loss from heat gradients.

Increasing the reflectivity of the crystal by changing the refractive index is one possible means to increase the optomechanical coupling and reduce the filter cavity circulating power as per equation 3 of the main text. The dependence of the optomechanical frequency shift versus BAW resonator refractive index is shown in figure \ref{fig:indexCoupling}. Coating the surface of the BAW resonator with one layer of $\lambda/4$ dielectric can further enhance the reflectivity and optomechanical coupling. For example, using a quarter wave layer of silicon nitride on both sides of the resonator raises the effective refractive index of the crystal to 2.54. Given the thickness ratio of coating to substrate (hundreds of nm versus 1 mm), there is a possibility that the mechanical Q-factor will not be degraded too much by the surface treatment, but this bears more detailed investigation in future work. 

It may be possible to design a BAW resonator that can support Brillouin scattering interactions at a high mechanical quality factor. We also note that future GW detectors may use $\lambda = 2$ $\mu$m optics, which would reduce the BAW resonator's Brillouin scattering frequency by half. This is important for reducing the effect of surface scattering losses that limit the mechanical Q-factor of longitudinal bulk acoustic mechanical modes in the GHz frequencies. These considerations will be useful for the possibility of designing custom BAW resonators for GW detector optomechanics.

\section*{Reducing the power requirement for optomechanical coupling}

\begin{figure}
    \centering
    \includegraphics[width=0.45\textwidth]{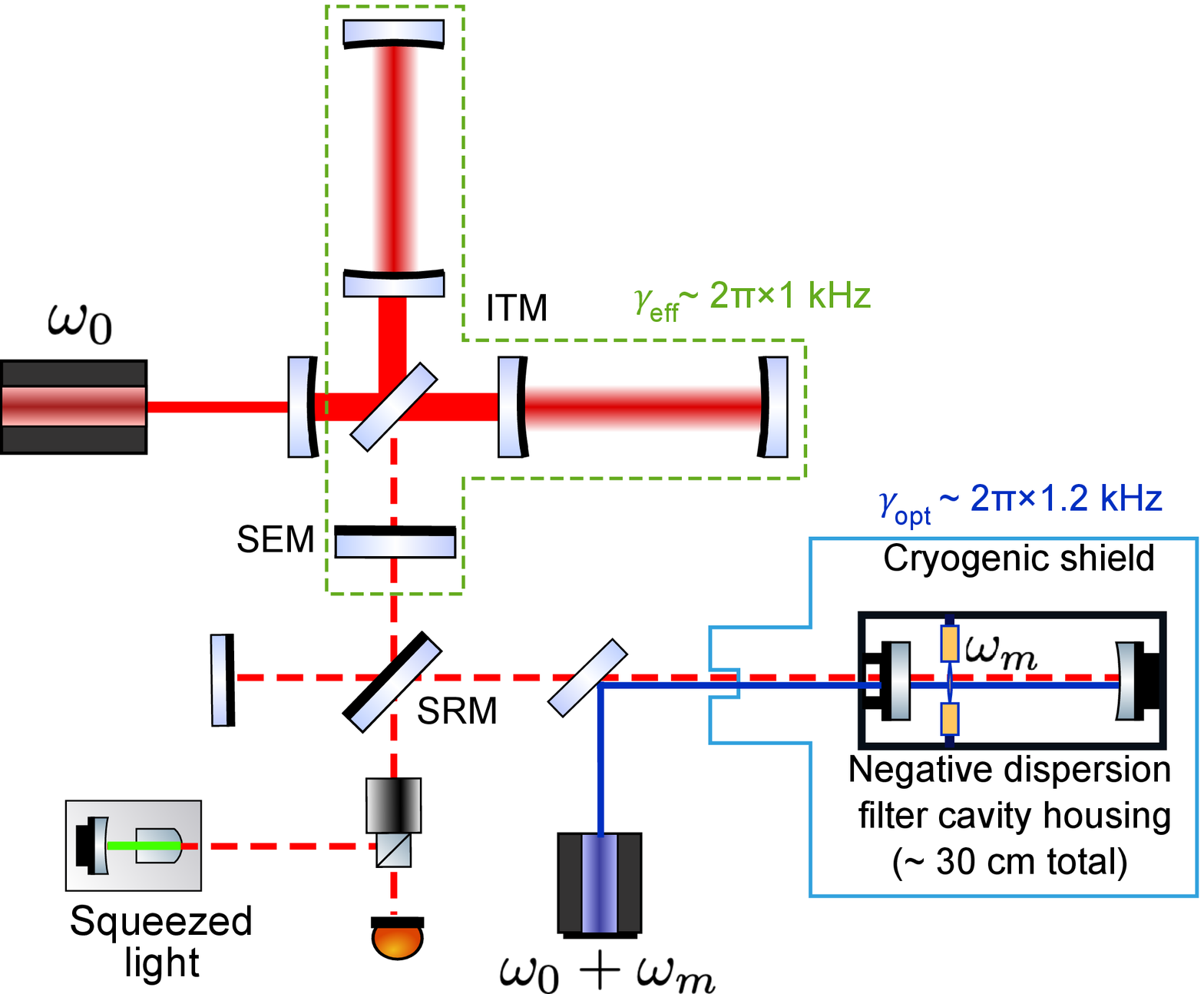}
    \caption{\textbf{Alternate WLSR configuration capable of using less filter cavity pumping power:} The interferometer is dual recycled, operating in resonant sideband extraction mode. Unlike figure 1 of the main text, the SEM is not perfectly impedance matched to the arm cavity. With respect to GW signals emerging out to the dark port, the interferometer can effectively be considered as a two mirror cavity with effective bandwidth $\gamma_{\rm eff}/(2\pi) = 1$ kHz.}
    \label{fig:optionB}
\end{figure}

\begin{figure}
    \centering
    \includegraphics[width=0.45\textwidth]{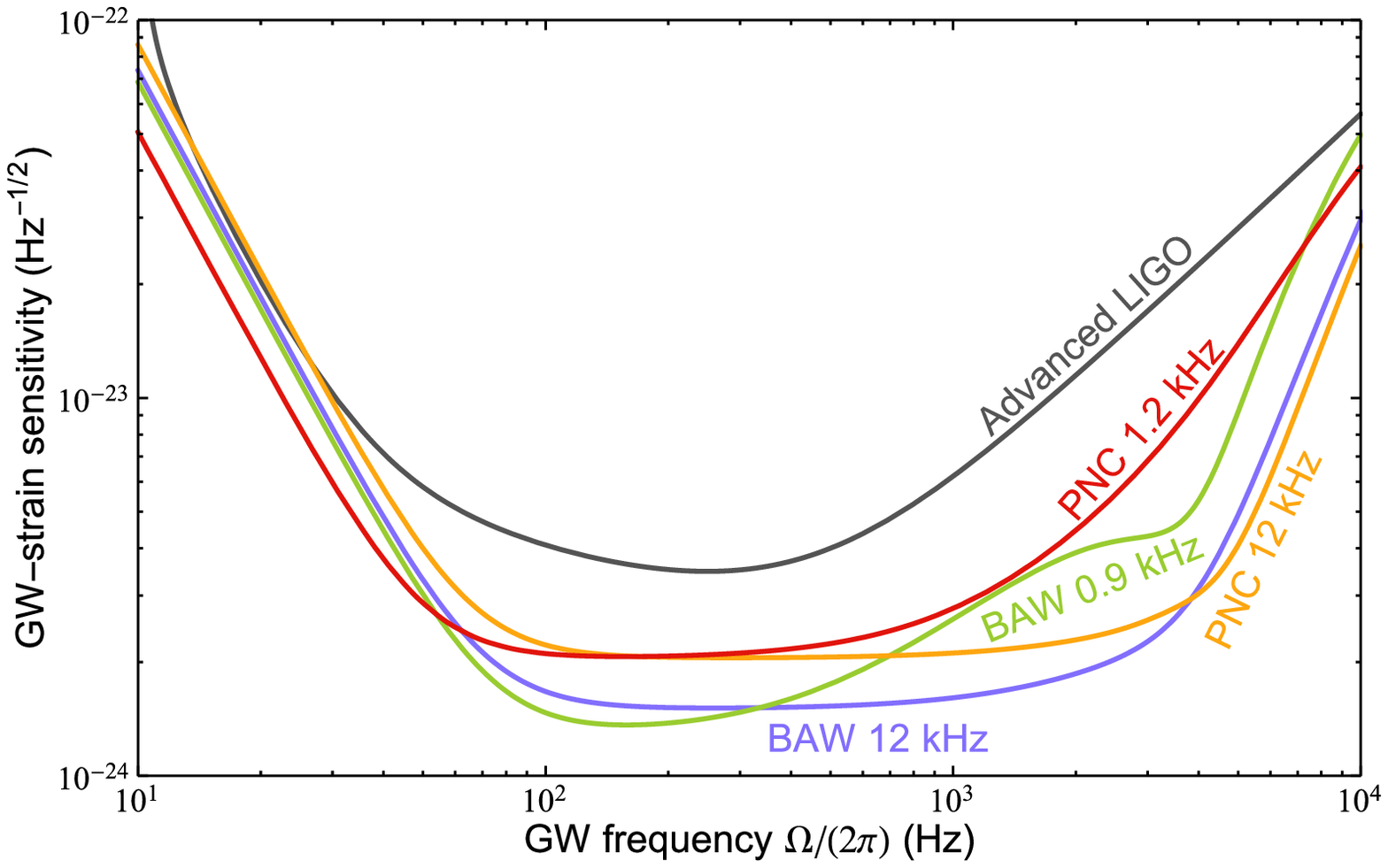}
    \caption{\textbf{Comparison of white light signal recycling with different optomechanical bandwidths:} Different WLSR schemes are denoted by the type of resonator and the optomechanical antidamping $\gamma_{\rm opt}/2\pi$. The schemes denoted by ``$12$ kHz'' use the layout shown in figure 1 of the main text, while the other WLSR schemes use the layout shown in figure \ref{fig:optionB}. We also note that the WLSR schemes here use 800 kW arm power and no application of frequency dependent squeezing.}
    \label{fig:optionAB}
\end{figure}

WLSR configurations can be made to use less optomechanical coupling $\gamma_{\rm opt}$, and thus less filter cavity power, at the expense of slightly lower bandwidth enhancement. This idea is useful for reducing the extremely large filter cavity power requirement of BAW resonators. The required layout is shown in figure \ref{fig:optionB}. The interferometer uses the dual-recycling Fabry-Perot Michelson configuration. Unlike the SEM in figure 1 of the main text, the SEM here is not perfectly impedance matched to the arms. Instead, the ITM and SEM are configured such that they form a compound mirror with effective arm cavity bandwidth of $\gamma_{\rm arm}/(2\pi) = 1$ kHz \cite{BuonnanoScaling}. This can be achieved with $T_{\rm ITM} = 0.033$ and $T_{\rm SEM} = 0.33$. 

The result of WLSR using reduced $\gamma_{\rm opt}$ is shown in figure \ref{fig:optionAB}. These are compared to the WLSR curves using $\gamma_{\rm opt}/(2\pi) = 12$ kHz applied to the configuration of figure 1 of the main text. The WLSR curves of figure \ref{fig:optionAB} use 800 kW arm cavity power, zero squeezed vacuum input and interferometer optical losses are comparable to the target loss of the A+ upgrade \cite{LIGOWP2019}. Filter cavity round trip losses are maintained near 20 ppm in order to bring their respective optical loss close to the quantum noise level at 1--5 kHz. Full noise budgets for the curves shown in figure \ref{fig:optionAB} are detailed in the following section.

\section*{Noise budget of WLSR interferometers}

\begin{table*}[]
    \centering
    \begin{tabular}{|c c c c c c|}\hline
     Parameter  & PNC & BAW & BAW ULL & PNC 2 km & PNC 10 km \\ \hline
     \textbf{Interferometer} & & & & & \\
                Circulating arm power $P$ (MW) & 4.0 & 4.0 & 4.0 & 4.0 & 4.0 \\
                Arm length $L_{\rm arm}$ (km) & 4.0 & 4.0 & 4.0 & 2.0 & 10.0 \\
            Observed squeezing (dB) & 10 & 10 & 10 & 10 & 10 \\
       SRM transmission $T_{\rm sr}$  & 0.025 & 0.01 & 0.01 & 0.01 & 0.1 \\
                Input test mass transmission $T_{\rm itm}$ & 0.04 & 0.04 & 0.04 & 0.04 & 0.04 \\
                Output to photodiode loss & 0.025 & 0.025 & 0.01& 0.025 & 0.025 \\
                 Total arm losses $\epsilon_{\rm arm}$ (ppm) &1200 & 1200 & 800 & 850 &3000 \\
                 ITM thermal compensation $\kappa_{\rm ITM}$ &70 & 70 & 90 & 70 & 70\\
                 Beamsplitter thermal comp. $\kappa_{\rm BS}$ & 10 & 10 & 10 & 10 & 10  \\
                 ITM absorption $\alpha_{\rm ITM}$ (ppm)&  0.25 & 0.25 & 0.25 & 0.25 & 0.25 \\
                 Beamsplitter absorption $\alpha_{\rm BS}$ (ppm)& 1.0 & 1.0 & 0.5 & 1.0 & 1.0 \\
               Signal extraction losses $\epsilon_{\rm se}$ (ppm)& 957  & 957  & 500 & 957 & 957 \\ \hline
               \textbf{Filter cavity} & & & & & \\
               OM coupling $\gamma_{\rm opt}/2\pi$ (kHz) & 11.5 & 11.5 & 11.5 & 23.4 & 4.34\\
                Filter cavity roundtrip loss $\epsilon_{f}$ (ppm)  & 10 &5 & 5 & 10 & 25  \\
                Filter cavity transmission $T_f$ (ppm)&300 & 250  &  500 & 200 & 750 \\
                Filter cavity length $L_f$ (m) & 0.05 &0.05 &0.05 & 0.05 & 0.05 \\
                 Thermal noise coupling $T/Q_m$ (K) & 1$\times$10$^{-9}$ & 5$\times$10$^{-10}$ &  6.5$\times$10$^{-11}$ & 1$\times$10$^{-9}$ & 1$\times$10$^{-9}$ \\ \hline
                \end{tabular}
    \caption{Properties of WLSR interferometer schemes that are based on future GW detector configurations. The first 3 columns refer to the key curves in figure 1 of the main text. Figure \ref{fig:lengthComparison} shows WLSR with different interferometer lengths, denoted in the latter two columns. Impedance matching between the arm cavity and signal recycling cavity is obtained by setting $T_{\rm SEM}$ equal to $T_{\rm ITM}$. An ultra low-loss (ULL) WLSR configuration using the BAW resonator is formulated based on the prospective $T/Q_m$ that can be obtained by decreasing the BAW crystal temperature to 1 K, as per figure \ref{fig:TLSQ}.}
    \label{tab:BasicAdvancedUltra}
\end{table*}

\begin{table*}[]
    \centering
    \begin{tabular}{|c c c c c c|}\hline
     Parameter & PNC 1.2 kHz &  BAW 0.9 kHz &PNC 12 kHz& BAW 12 kHz &  PNC 200 kW  \\ \hline
     \textbf{Interferometer} & & & & & \\
                Circulating arm power $P$ (MW) & 0.80 & 0.80 & 0.80 & 0.80 & 0.20 \\
                Arm length $L_{\rm arm}$ (km) & 4.0 & 4.0 & 4.0 & 4.0 & 4.0 \\
            Observed squeezing (dB) & 0 & 0 & 0 & 0 & 0 \\
       Signal recycling transmission $T_{\rm sr}$  & 0.2 & 0.1 & 0.01 & 0.01 & 0.01 \\
                Input test mass transmission $T_{\rm itm}$ & 0.033 & 0.033& 0.033 & 0.033 & 0.033 \\
                 Output to photodiode loss & 0.05& 0.05& 0.05 & 0.05 & 0.05 \\
                 Total arm losses $\epsilon_{\rm arm}$ (ppm) &12400&7400&7400 &  7400 & 9400 \\
                 ITM thermal compensation $\kappa_{\rm ITM}$ &30& 30& 60 & 60 & 30 \\
                 Beamsplitter thermal comp. $\kappa_{\rm BS}$ & 1& 1 & 2.5 & 2.5 & 1 \\
                 ITM absorption $\alpha_{\rm ITM}$ (ppm)& 0.5& 0.5& 0.5 & 0.5 & 0.5 \\
                 Beamsplitter absorption $\alpha_{\rm BS}$ (ppm)& 1.0 & 1.0 & 1.0 & 1.0 & 1.0 \\
               Signal extraction losses $\epsilon_{\rm se}$ (ppm)& 1250&  1250&  257 &  257  &  78\\ \hline
               \textbf{Filter cavity} & & & & &\\
               OM coupling $\gamma_{\rm opt}/2\pi$ (kHz) & 1.20 & 0.90 & 11.5 & 11.5 & 11.5 \\
                Filter cavity roundtrip loss $\epsilon_{f}$ (ppm) & 20 &  15 &20  & 15 &40 \\
                Filter cavity transmission $T_f$ (ppm)& 250 & 300 & 250 &150 & 400 \\
                Filter cavity length $L_f$ (m) & 0.05&0.05 &0.05 &0.05 &0.05 \\
                 Thermal noise coupling $T/Q_m$ (K) & 3$\times$10$^{-9}$ & 5$\times$10$^{-10}$ & 3$\times$10$^{-9}$ & 5$\times$10$^{-10}$  & 4$\times$10$^{-9}$ \\ \hline
                \end{tabular}
    \caption{Properties of WLSR interferometer schemes based on current and near-future GW detector configurations. The first four columns are used for the key curves in figure \ref{fig:optionAB}, where the number refers to the amount of optomechanical damping. Also shown are the parameters for a configuration using 200 kW arm cavity power, seen in figure \ref{fig:armPowerComparison}. Note that that temperature of the PNC resonator is raised compared to that shown in figure 1 of the main text.}
    \label{tab:Baselines}
\end{table*}

\begin{figure}
    \centering
    \includegraphics[width=0.45\textwidth]{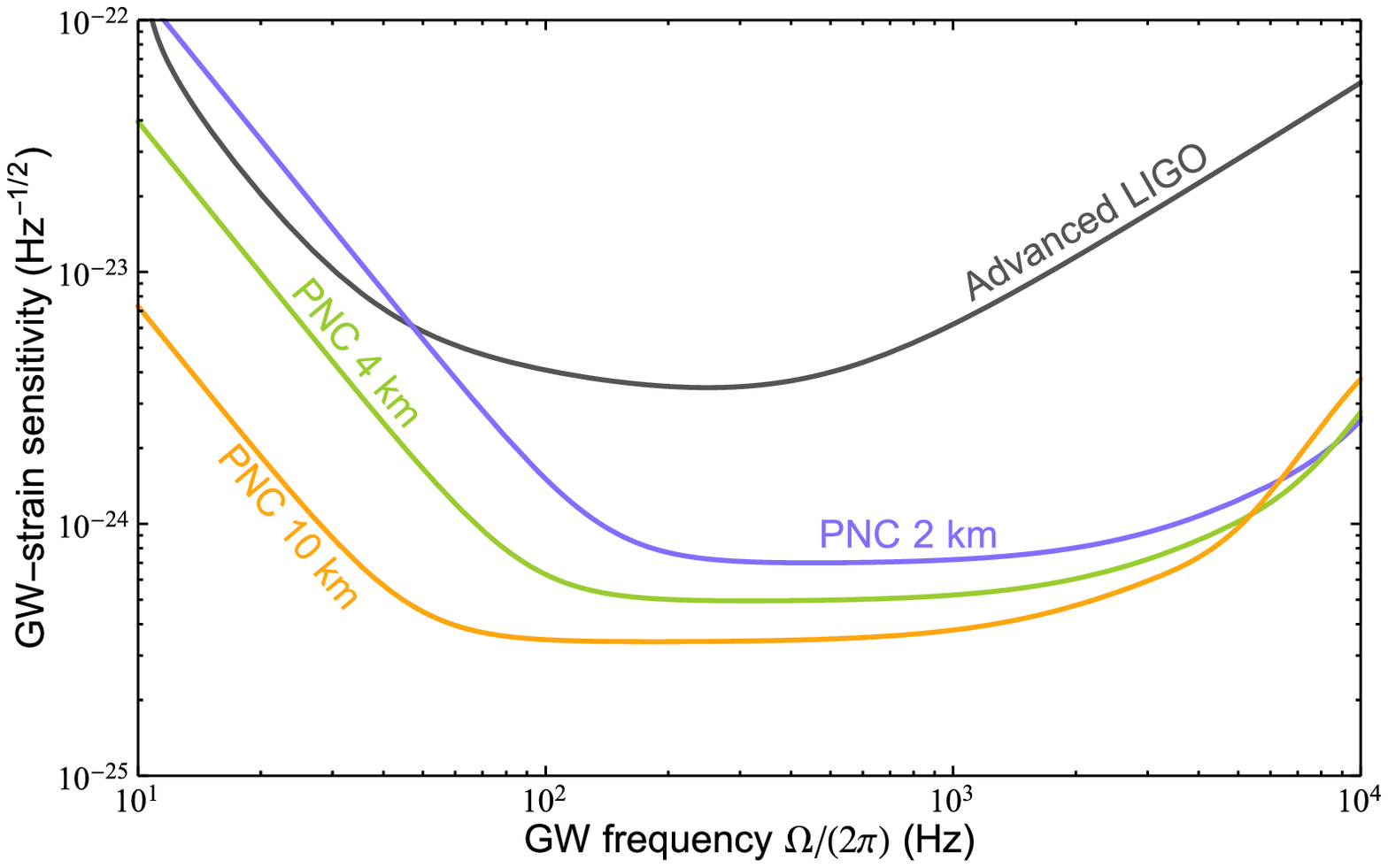}
    \caption{\textbf{WLSR vs interferometer arm lengths:} Comparison of WLSR across different interferometer arm lengths, using ``PNC'' loss parameters shown in table \ref{tab:BasicAdvancedUltra}. The 4 km curve is identical to the PNC result shown in figure 1 of the main text. All WLSR curves operate at 4 MW arm cavity circulating power and use 10 dB frequency dependent squeezing.}
    \label{fig:lengthComparison}
\end{figure}

\begin{figure}
    \centering
    \includegraphics[width=0.45\textwidth]{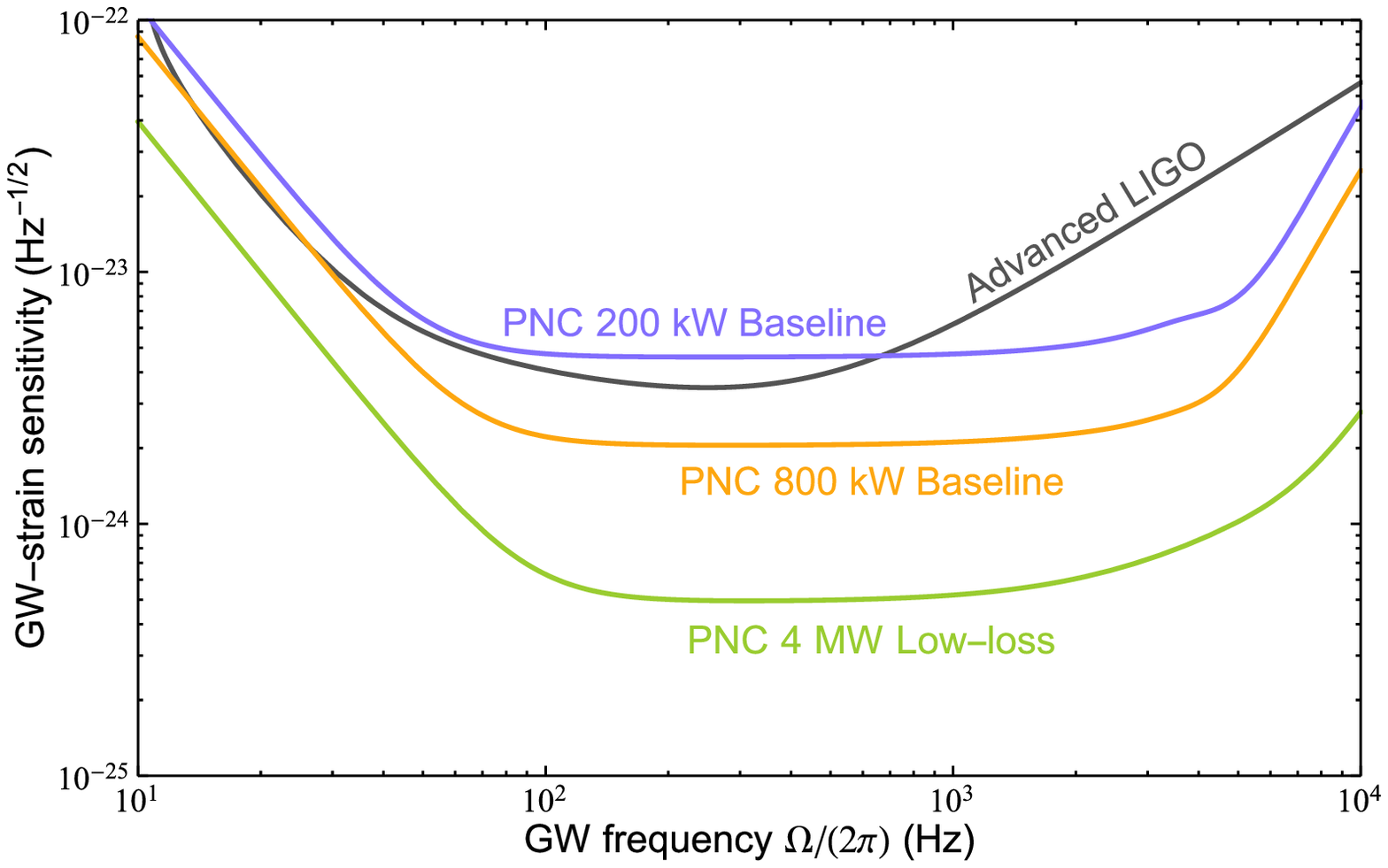}
    \caption{\textbf{WLSR at different levels of arm cavity power:} Comparison of WLSR across different interferometer arm powers, where `Baseline'' curves use optical loss parameters shown in the ``PNC 12 kHz'' and ``PNC 200 kW'' columns of table \ref{tab:Baselines}, while ``4 MW Low-loss'' is identical to the PNC curve shown in figure 1 of the main text.}
    \label{fig:armPowerComparison}
\end{figure}

\begin{figure*}[!ht]
    \centering
    \includegraphics[width=0.95\textwidth]{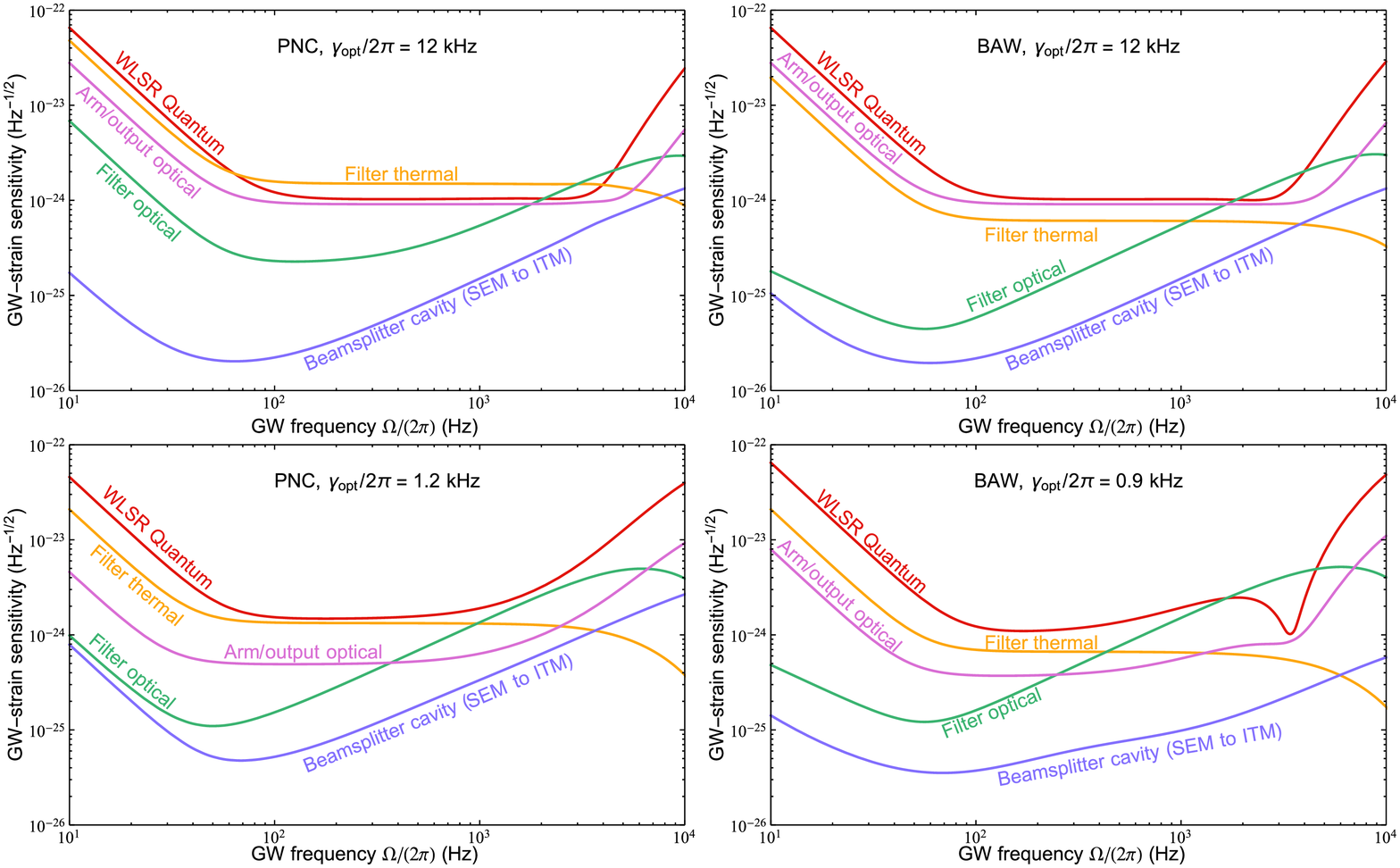}
    \caption{\textbf{Noise budget of various types of WLSR interferometers}: Breakdown of noise sources for WLSR interferometers, with totals shown as per figure \ref{fig:optionAB}. The title of each plot denotes the type of filter cavity mechanical resonator and the optomechanical anti-damping. The different noise sources are as follows: \textbf{WLSR Quantum:} Vacuum noise at the GW sideband frequency $\Omega$, referred to the output of the detector. No squeezing is applied in any of these plots. \textbf{Filter thermal:} Thermal noise introduced by the mechanical resonator inside the filter cavity. \textbf{Filter optical:} Optical loss introduced inside the filter cavity. \textbf{Beamsplitter cavity:} Power dependent optical losses that occur inside the beamsplitter cavity of the interferometer due to wavefront distortion. \textbf{Arm/output optical:} Sum of power independent optical losses that occur in the interferometer arm cavity, signal recycling cavity and throughout the output train of the detector's optics.}
    \label{fig:allBudgets}
\end{figure*}

Here we elaborate on further findings from the WLSR interferometer design framework outlined in the main text and Methods.

The WLSR sensitivity curves shown in figure 1 of the main text use a configuration loosely based on future GW detectors such as Einstein Telescope. The key features of these detectors with respect to this paper are the application of 10 dB frequency dependent squeezing to reduce quantum noise across the entire GW band, arm cavity power of several MW, improved thermal compensation of test mass distortion, reduced Brownian noise of test mass optical coatings and extremely low optical losses in the interferometer. Interferometer test mass coating thermal noise is unlikely to be a significant contributor in the high frequency band of interest. Relevant parameters used in the calculation of figure 1 of the main text are shown in table \ref{tab:BasicAdvancedUltra}.

The sensitivity of WLSR interferometers versus arm length are shown in figure \ref{fig:lengthComparison}, with parameters given in table \ref{tab:BasicAdvancedUltra}. The negative dispersion effect is approximately linear for $\Omega \ll \gamma_{\rm opt}$, as per equation \ref{eq:negativeDispersion} - since the required $\gamma_{\rm opt}$ is inversely dependent upon arm length, longer interferometers begin to exhibit non-linear filter cavity phase at lower $\Omega$. Longer interferometers also have a stricter thermal noise requirement from the filter cavity. However, their optical loss level is lower relative to the quantum noise floor.  In particular, the filter cavity optical loss can be reduced by increasing the filter cavity bandwidth, which reduces its fractional loss contribution. As per equation 3 in the main text, the filter cavity power requirement decreases with arm length, allowing the filter bandwidth to be increased to meet a certain pumping power target. It is seen that the 10 km WLSR interferometer can maintain superior sensitivity to a 4 km interferometers up to a frequency of 5 kHz, even given higher levels of $\epsilon_{\rm arm}$ and $\epsilon_{f}$ as per table \ref{tab:BasicAdvancedUltra}. In addition, the lower $\gamma_{\rm opt}$ requirement reduces the filter cavity circulating power, or alternately allows us to maintain a higher bandwidth filter cavity at the same level of pumping power. This allows us greater flexibility in WLSR design, and as such the technology will be useful for long-arm future GW detectors such as Einstein Telescope and Cosmic Explorer.

The WLSR sensitivity curves shown in figure \ref{fig:optionAB} use a configuration loosely based on current and near-future GW detectors such as Advanced LIGO and A+. The arm cavity maintains 800 kW of optical power, no frequency dependent squeezing is applied, and interferometer optical losses are approximately 1\%. Relevant parameters used in the calculation of figure \ref{fig:optionAB} are shown in table \ref{tab:Baselines}. Also shown is a set of parameters related to an interferometer using 200 kW of arm cavity power, which is approximately the current level that has been achieved in Advanced LIGO. Figure \ref{fig:armPowerComparison} shows a comparison of three different WLSR configurations - one using 200 kW arm power, one using 800 kW arm power and one using 4 MW arm power with 10 dB FDS. It is seen that the 200 kW WLSR interferometer can reach a peak sensitivity comparable to that of the nominal Advanced LIGO design at 800 kW arm power, but with bandwidth that extends to the NS frequencies at 1--5 kHz.

A breakdown of noise sources for a set of baseline 4 km WLSR interferometers is shown in figure \ref{fig:allBudgets}. These correspond to the respective total noise curves shown in figure \ref{fig:optionAB}. Loss parameters are shown in the relevant columns of table \ref{tab:Baselines} corresponding to the type of resonator and optomechanical antidamping value. It is seen that the baseline WLSR interferometer with $\gamma_{\rm opt}/(2\pi) = 12$ kHz achieves sensitivity of approximately $1.5\times10^{-24}$ Hz$^{-1/2}$ for frequencies from 150 Hz to 4 kHz. For the case of the PNC, the resulting sensitivity is limited by thermal noise from the resonator. Arm/SRC losses at the level of 1\% are comparable to the quantum noise floor. For both the PNC and BAW resonators, the filter cavity optical loss is significant in the band 1--5 kHz. This is due to the fact that the finesse of the filter cavity must be high to reduce the circulating power required, but this increases the fractional optical loss. In addition, the cases with lower $\gamma_{\rm opt}$ also display high filter cavity optical loss in the NS band. This is because the leading coefficient of noise terms in the filter cavity Heisenberg equations of motion have $\gamma_{\rm opt} - i\Omega$ in the denominator, causing their effects to become significant as $\Omega \rightarrow \gamma_{\rm opt}$.

A high-frequency detector scheme using Advanced LIGO-style dual recycling topology was analysed by Martynov \textit{et al.} \cite{DenisHF}, which exploits optical sloshing between the arm cavity and SRC. The optical resonance and bandwidth of the sloshing interaction between the arm cavity and SRC is mediated by transmissivities $T_{\rm ITM}$, $T_{\rm sr}$ and cavity lengths $L_{\rm arm}$, $L_{\rm src}$ of the arm cavity and SRC. Tuning the signal recycling parameters allows for the sloshing interaction to resonantly enhance signals in the neutron star frequency band. A comparison of a dual recycling sloshing interferometer scheme with low-loss WLSR is shown in figure 1 of the main text. The noise curve for the sloshing interferometer accounts for quantum noise as well as resonantly enhanced thermal lensing in the beamsplitter cavity. The ``sloshing SR'' and ``PNC'' curves use identical thermal lensing compensation parameters shown in the ``PNC'' column of table \ref{tab:BasicAdvancedUltra}. It can be seen that WLSR is capable of achieving better sensitivity over a broader band of neutron star frequencies, given the same level of squeezing, arm length and thermal lensing compensation. The WLSR curve shown in figure 1 of the main text also additionally accounts for arm/SRC losses, whereas the sloshing SRC curve in the same figure disregards arm/SRC losses and thus its relative sensitivity is slightly overestimated.

The two important internal loss mechanisms for the BAW resonator are the Landau-Rumer process, and TLS interaction. Extrapolating the loss vs temperature into the TLS limit indicates that $T/Q_m = 6.5 \times 10^{-11}$ K may be possible at 1 K, as per figure \ref{fig:TLSQ}. This potential improvement is used for an ultra low-loss configuration illustrated in figure 1 of the main text, and detailed in the ``BAW ULL'' column of table \ref{tab:BasicAdvancedUltra}. Compared to the $T/Q_m$ value used for thermal noise coupling calculations in figure \ref{fig:allBudgets}, the extrapolated thermal noise value represents a factor of 7.5 improvement in thermal noise coupling from the filter cavity to the interferometer, and a $\sqrt{7.5}$ improvement in the thermal noise contribution to the total strain sensitivity amplitude. The ultra low-loss 4 km WLSR interferometer is capable of reaching $h<5\times 10^{-25}$ Hz$^{1/2}$ in the neutron star frequency band.

\section*{Absorption heating of mechanical resonators}

\begin{figure}
    \centering
    \includegraphics[width = 0.45\textwidth]{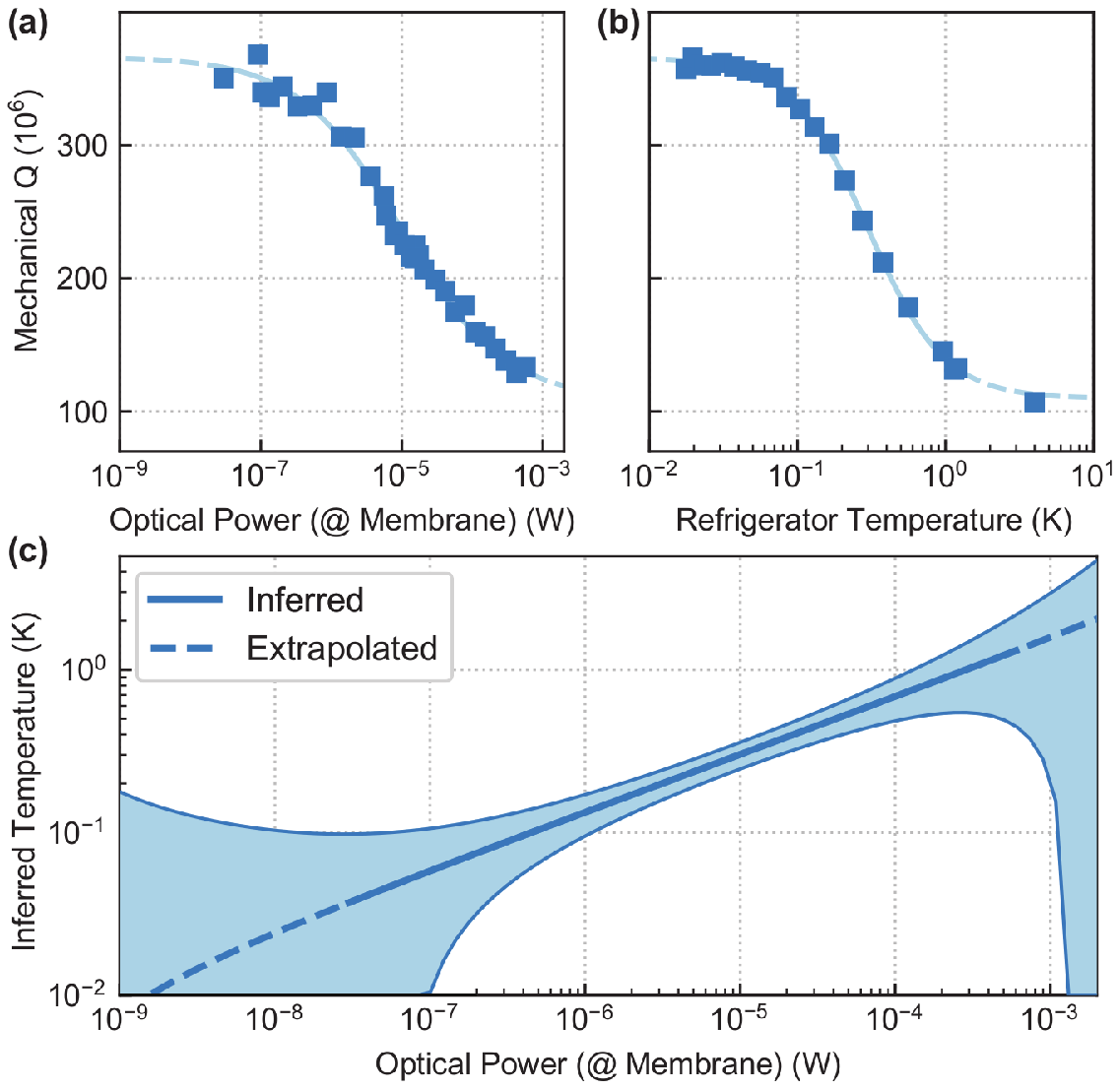}
    \caption{\textbf{Low-temperature mechanical properties of a phononically shielded SiN membrane resonator under optical illumination:} (a) Mechanical quality factor vs optical power traversing the membrane held at a refrigerator temperature of 15 mK. (b) Mechanical quality factor (probed with low optical power) vs refrigerator temperature. (c) Taking the mechanical quality factor as a proxy for the membrane temperature, we infer the effect of absorption heating. To this end, the generic polynomial model shown as line in panel (b) is inverted, and applied to the quality factors shown in panel (a). }
    \label{fig:measuredPNCTemperature}
\end{figure}

\begin{figure}
    \centering
    \includegraphics[width=0.45\textwidth]{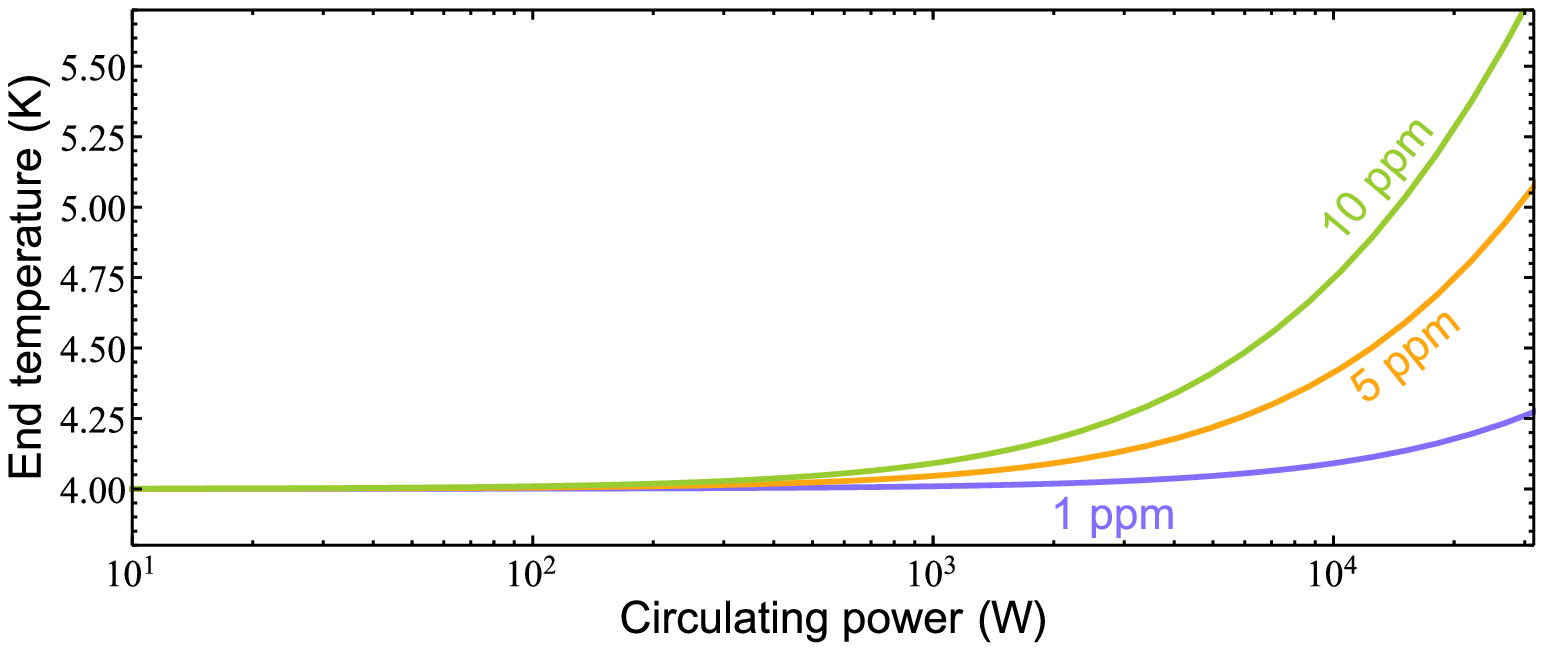}
    \caption{\textbf{One-dimensional estimate of equilibrium temperature of the quartz BAW resonator in the negative dispersion filter:} Various levels of optical absorption are shown, though single-crystal quartz BAW resonators can be reasonably expected to have less than 1 ppm absorption of 1064 nm light. The contacts to the BAW resonator are maintained at 4 K. Even at the high powers required for WLSR with BAW resonators, a first estimate indicates that maintaining the resonator at cryogenic temperature is plausible.}
    \label{fig:quartzTemperature}
\end{figure}

Heating of the resonators used in the negative dispersion filter may be a concern. In the case of the PNC it is due to the thermal resistance of the geometry, namely the thinness of the resonator and phononic shield. In the case of the quartz BAW resonator it is due to the high laser power incident on the surface that is necessitated by the filter cavity pumping requirement. 

Measured low temperature mechanical properties of a silicon nitride PNC resonator are shown in figure \ref{fig:measuredPNCTemperature}. It is seen that it is possible to maintain the temperature of a the sample close to 5 K for a few mW of incident laser power. Note that the sample in question differs from that referred to throughout this paper, in that the resonator is 60 nm thick rather than 20 nm thick, and $\sim 800$ nm laser light was used, which has slightly higher optical absorption in Si$_3$N$_4$. Thin resonators are required for higher quality factors, and also have less optical absorption. However, thicker resonators also have less thermal resistance.

Silicon nitride membranes have been shown to have low optical absorption of near-infrared light. Wilson has suggested an upper limit to the refractive index imaginary component of $\mathrm{Im}(n_m)\leq 0.8\times 10^{-5}$, corresponding to a power absorption of $\lesssim$ 4.5 ppm for a 20 nm thick membrane \cite{WilsonThesis}. Sankey \textit{et al.} report even lower absorption of $\mathrm{Im}(n_m)\leq 1.5\times 10^{-6}$ for Si$_3$N$_4$ with 1064 nm light \cite{SankeyNonlinear}. In addition, Peterson \textit{et al.} measured optomechanical damping of the motion of a 40 nm thick Si$_3$N$_4$ membrane resonator, and suggested that material absorption was not a dominant contributor to the bath temperature of 360 mK, even with input power of 5 $\mu$W into a cavity with finesse 57,000 \cite{RegalHeating}. These measurements of low absorption with membrane resonators are encouraging, however, due to the nature of manufacturing microresonators, we emphasise that the projections of absorption heating are meant as a guide only.

For the BAW resonator, a 1-dimensional order of magnitude estimate of resonator heating is obtained using the following conduction law:

\begin{equation}
    P_{\rm conduct} = P_{\rm absorb} = -\sigma \kappa(T) \dfrac{d T(z)}{d z}
\end{equation}
\noindent
where $P_{\rm conduct}$ is the power conducted through a channel of cross sectional area $\sigma$ and length dimension $z$ in the direction of heat conduction. $P_{\rm conduct}$ is assumed to be equal to the absorbed optical power $P_{\rm absorb}$. Assuming that, at cryogenic temperatures, the thermal conductivity can be approximated by $\kappa = \kappa_0 T^n$ and integrating with respect to $z$ gives:

\begin{equation}
    T_{\rm equilib} = \left(\dfrac{P_{\rm absorb}(n+1)l_{\rm lim}}{\kappa_0 \sigma}+T_{\rm external}^{n+1}\right)^{1/(n+1)}
\end{equation}

\noindent
where $l_{\rm lim}$ is the length of the limiting component of thermal resistivity, $T_{\rm external}$ is the temperature of cryogenics for which we nominally maintain the resonator, and $T_{\rm equilib}$ is the equilibrium temperature of the resonator assuming that heat can escape via conduction. For the BAW resonator we look at conduction through a channel with length of $l_{\rm lim} = 15$ mm and cross sectional area $\sigma =5$ mm$^2$. External cryogenics are maintained at 4 K. The cryogenic temperature dependent thermal conductivity for quartz is obtained from values measured by Hofacker and Lohneysen \cite{HofackerQuartzConductivity} and below 5 K is approximately equal to $\kappa \sim 10~T^{2.5}$.

The equilibrium temperatures of the BAW resonator is shown in figure \ref{fig:quartzTemperature} for three different values of optical absorption. In the approximation of 1-dimensional conduction, maintaining the resonator at less than 1 K temperature difference from the environment is plausible, but would require optical absorption of less than 5 ppm. Fused quartz used in GW detector optics has an absorption coefficient of less than 1 ppm/cm at 1064 nm due to high purity \cite{QuartzAbsorption}, which contributes to its extremely high damage threshold.

\bibliography{bulkRefs}